\documentclass[preprint]{elsarticle}

\usepackage[utf8]{inputenc}
\usepackage[T1]{fontenc}
\usepackage{lmodern}
\usepackage{amsmath,amssymb}
\usepackage{bbm} 
\usepackage[top=3cm,left=3cm,right=3cm,bottom=4cm,footskip=3em]{geometry}
\usepackage{booktabs}
\usepackage{listings}
\usepackage{scalefnt}
\usepackage[usenames]{xcolor}
\usepackage[
  pdftitle={Two-loop Prediction of the Anomalous Magnetic
    Moment of the Muon in the Two-Higgs Doublet Model with GM2Calc 2},
  pdfauthor={
    Peter Athron,
    Csaba Balazs,
    Adriano Cherchiglia,
    Douglas HJ Jacob,
    Dominik Stöckinger,
    Hyejung Stöckinger-Kim,
    Alexander Voigt},
  pdfkeywords={2HDM, anomalous magnetic moment},
  bookmarks=true,
  linktocpage,
  colorlinks=true,
  allbordercolors=white,
  allcolors=fscolor]{hyperref}
\usepackage{orcidlink} 
\usepackage{textcomp}

\biboptions{sort&compress}

\definecolor{red}{rgb}{0.8,0,0}
\definecolor{green}{rgb}{0,0.5,0}
\definecolor{blue}{rgb}{0,0,0.9}

\newcommand{\FS}{\texttt{Flex\-ib\-le\-SU\-SY}}
\newcommand{\FD}{\texttt{Flex\-ib\-le\-De\-cay}}
\newcommand{\GMTCalc}{\texttt{GM2Calc}}
\newcommand{\THDMC}{\texttt{2HDMC}}
\newcommand{\THDECAY}{\texttt{2HDECAY}}
\newcommand{\ScannerS}{\texttt{ScannerS}}
\newcommand{\PROPHECYfourF}{\texttt{PROPHECY4F}}
\newcommand{\SARAH}{\texttt{SARAH}}
\newcommand{\spheno}{\texttt{SPheno}}
\newcommand{\HiggsBounds}{\texttt{HiggsBounds}}
\newcommand{\HiggsSignals}{\texttt{H{\scalefont{.9}IGGS}S{\scalefont{.9}IGNALS}}}
\newcommand{\GAMBIT}{\texttt{GAMBIT}}
\newcommand{\mathematica}{\texttt{Ma\-the\-ma\-ti\-ca}}
\newcommand{\python}{\texttt{Py\-thon}}
\newcommand{\aem}{\ensuremath{\alpha_{\text{em}}}}
\newcommand{\amu}{\ensuremath{a_\mu}}
\newcommand{\amuB}{\ensuremath{\amu^{\text{B}}}}
\newcommand{\amuF}{\ensuremath{\amu^{\text{F}}}}
\newcommand{\amuBSM}{\ensuremath{\amu^{\BSM}}}
\newcommand{\Damu}{\ensuremath{\delta\amu}} 
\newcommand{\ol}{\ensuremath{{1\ell}}}
\newcommand{\tl}{\ensuremath{{2\ell}}}
\newcommand{\thl}{\ensuremath{{3\ell}}}
\newcommand{\hc}{\text{h.\,c.}}
\newcommand{\order}{\mathcal{O}}
\DeclareMathOperator{\dilog}{Li_2}
\newcommand{\F}{\mathcal{F}} 
\newcommand{\PiL}{\Pi^0} 
\newcommand{\rhoL}{\rho^0} 
\newcommand{\PL}{P_{\scalefont{.9}\text{L}}}
\newcommand{\PR}{P_{\scalefont{.9}\text{R}}}
\newcommand{\real}{\Re\mathfrak{e}}
\newcommand{\VCKM}{V_{\scalefont{.9}\text{CKM}}}
\newcommand{\SM}{{\scalefont{.9}\text{SM}}}
\newcommand{\BSM}{{\scalefont{.9}\text{BSM}}}
\newcommand{\MSSM}{{\scalefont{.9}\text{MSSM}}}
\newcommand{\THDM}{{\scalefont{.9}\text{2HDM}}}
\newcommand{\FATHDM}{{\scalefont{.9}\text{FA2HDM}}}
\newcommand{\MS}{{\overline{\scalefont{.9}\text{MS}}}}
\newcommand{\hSM}{\ensuremath{{h_{\SM}}}}
\newcommand{\SU}{\ensuremath{\text{SU}}}
\newcommand{\id}{\ensuremath{\mathbbm{1}}}
\newcommand{\unit}[1]{\,\text{#1}}
\newcommand{\GeV}{\unit{GeV}}
\DeclareMathOperator{\diag}{diag}
\renewcommand{\imath}{\text{i}}

\newcommand{\figref}[1]{\figurename~\ref{#1}}
\newcommand{\secref}[1]{Section~\ref{#1}}
\newcommand{\tabref}[1]{\tablename~\ref{#1}}
\newcommand{\vac}{v}
\newcommand{\Mh}{h}
\newcommand{\MH}{H}

\newcommand{\MHpm}{H^{\pm}}

\newcommand{\MMh}{m_{\Mh}}
\newcommand{\MMH}{m_{\MH}}

\newcommand{\MMHpm}{m_{\MHpm}}

\begin{document}
\begin{frontmatter}
  \title{\Large \bf Two-loop Prediction of the Anomalous Magnetic
    Moment of the Muon in the Two-Higgs Doublet Model with GM2Calc 2}
  \author[nanjing,monash]{Peter Athron\orcidlink{0000-0003-2966-5914}}
  \author[monash]{Csaba Balazs\orcidlink{0000-0001-7154-1726}}
  \author[standre]{Adriano Cherchiglia\orcidlink{0000-0002-2457-1671}}
  \author[monash]{Douglas Jacob\orcidlink{0000-0002-8950-2853}\corref{cor1}}
  \ead{douglas.jacob@monash.edu}
  \cortext[cor1]{Corresponding author}
  \author[dresden]{Dominik Stöckinger}
  \author[dresden]{Hyejung Stöckinger-Kim}
  \author[flensburg]{Alexander Voigt\orcidlink{0000-0001-8963-6512}}
  \address[nanjing]{Department  of  Physics  and  Institute  of  Theoretical  Physics,  Nanjing  Normal  University, Nanjing, Jiangsu 210023, China}
  \address[monash]{ARC Centre of Excellence for Particle Physics at
    the Terascale, School of Physics, Monash University, Melbourne,
    Victoria 3800, Australia}
  \address[standre]{Centro de Ci\^ecias Naturais e Humanas, Universidade Federal do ABC, Santo Andr\'e, Brazil}
  \address[dresden]{Institut für Kern- und Teilchenphysik,
    TU Dresden, Zellescher Weg 19, 01069 Dresden, Germany}
  \address[flensburg]{Fachbereich Energie und Biotechnologie, Hochschule Flensburg, Kanzleistraße 91--93, 24943 Flensburg, Germany}

  \begin{abstract}
  We present an extension of the \GMTCalc\ software to calculate the
  muon anomalous magnetic moment ($\amuBSM$) in the
  Two-Higgs Doublet Model.  The Two-Higgs Doublet Model is one of the
  simplest and most popular extensions of the Standard Model.  It is
  one of the few single field extensions that can give large
  contributions to $\amuBSM$. It is essential to include two-loop
  corrections to explain the long standing discrepancy between the
  Standard Model prediction and the experimental measurement in the
  Two-Higgs Doublet Model.
  The new version \GMTCalc\ 2 implements the state of the art two-loop
  calculation for the general, flavour violating Two-Higgs Doublet
  Model as well as for the flavour aligned Two-Higgs Doublet Model and
  the type I, II, X and Y flavour conserving variants.
  Input parameters
  can be provided in either the gauge basis or the mass basis, and
  we provide an easy to use SLHA-like command-line interface to
  specify these.  Using this interface users may also select between Two-Higgs Doublet Model types and choose which contributions
  to apply.  In addition, \GMTCalc\ 2 also provides interfaces in C++, C, \python\
  and \mathematica, to make it easy to interface with other codes.
\end{abstract}

\end{frontmatter}

\clearpage
\tableofcontents
\clearpage

 \section{Introduction}

 The anomalous magnetic moment of the muon, $\amu$, is one of the most
 important observables in particle physics.  Its importance stems from
 the fact that it is one of the most precisely measured physical 
 quantities and that it is very sensitive to new
 physics as well as the strong, weak and electromagnetic interactions
 of the Standard Model (\SM).  The significance of $\amu$ has increased further
 in the wake of the recent release of the first results from the
 Fermilab Muon $g-2$ experiment \cite{PhysRevLett.126.141801}.  This collaboration
 found, in combination with the results from the Brookhaven National
 Laboratory \cite{Tanabashi2018}, an experimental value of
\begin{equation} \label{eqn:Gm2Exp}
	\amu^\text{Exp} = (11659206.1\pm4.1)\times10^{-10}.
\end{equation}
The \SM\ prediction from the Muon $g-2$ Theory Initiative White Paper \cite{aoyama2020anomalous} is
\begin{equation} \label{eqn:Gm2SM}
	\amu^{\SM} = (11659181.0\pm4.3)\times10^{-10},
\end{equation}
which combines quantum electrodynamic contributions \cite{Aoyama:2012wk,Aoyama:2019ryr}, 
electroweak contributions \cite{Czarnecki:2002nt,Gnendiger:2013pva}, 
hadronic vacuum-polarization \cite{Davier:2017zfy,Keshavarzi:2018mgv,Colangelo:2018mtw,Hoferichter:2019gzf,Davier:2019can,Keshavarzi:2019abf,Kurz:2014wya}
and light-by-light contributions \cite{Melnikov:2003xd,Masjuan:2017tvw,Colangelo:2017fiz,Hoferichter:2018kwz,Gerardin:2019vio,Bijnens:2019ghy,Colangelo:2019uex,Pauk:2014rta,Danilkin:2016hnh,Jegerlehner:2017gek,Knecht:2018sci,Eichmann:2019bqf,Roig:2019reh,Blum:2019ugy,Colangelo:2014qya}.\footnote{%
  As discussed extensively in the Theory Initiative White Paper
  \cite{aoyama2020anomalous}, the proposed \SM\ prediction does not use
  lattice gauge theory evaluations of the hadronic vacuum
  polarization. The lattice world average evaluated in
  Ref.\ \cite{aoyama2020anomalous}, based on \cite{Chakraborty:2017tqp,Borsanyi:2017zdw,
Blum:2018mom,Giusti:2019xct,Shintani:2019wai,Davies:2019efs,Gerardin:2019rua,Aubin:2019usy,Giusti:2019hkz}, is compatible
  with the data-based result 
\cite{Davier:2017zfy,Keshavarzi:2018mgv,Colangelo:2018mtw,Hoferichter:2019gzf,Davier:2019can,Keshavarzi:2019abf,Kurz:2014wya}, has a higher
  central value and larger uncertainty. However more recent lattice results
  are obtained in Refs.\ \cite{Borsanyi:2020mff,Lehner:2020crt}, and
  in particular Ref.\ \cite{Borsanyi:2020mff} obtains a result with smaller
  uncertainty, which would shift the \SM\ prediction for $\amu$ closer to the
  experimental value. Scrutiny of these results is
  ongoing (see e.g.\ Ref.\ \cite{Colangelo:2020lcg}) and further
  progress can be expected.\label{footnoteLQCD}
  }
The experimental measurement differs from the \SM\ prediction by $4.2\sigma$ which suggests a beyond the Standard
Model (\BSM) contribution of
\begin{equation}
	\amuBSM = (25.1\pm5.9)\times10^{-10}.
        \label{eq:amuBSM}
\end{equation} 
The high precision in both the experimental measurement and the \SM\ theoretical prediction requires \BSM\ contributions to be known with a similar level of accuracy. 

The Two-Higgs Doublet Model (\THDM) is one of the simplest and most widely studied extensions of the \SM\ 
(see e.g.\ Ref.\ \cite{Branco:2011iw} for a review or the relevant chapters of Ref.\ \cite{Gunion:1989we}). 
Although the \THDM\ has been studied for many decades and the only Higgs boson discovered is closely \SM-like \cite{ATLAS:2012yve,CMS:2012qbp},
interest in this model has not waned, and recent progress has been
achieved e.g.\ on LHC interpretations 
\cite{Haber:2015pua,Baglio:2014nea,Eberhardt:2013uba,Chowdhury:2015yja,Chowdhury:2017aav}, 
B-physics \cite{Misiak:2017bgg,Hu:2016gpe,Li:2014fea,Enomoto:2015wbn,Arnan:2017lxi}, 
theoretical constraints \cite{Barroso:2013awa,BhupalDev:2014bir,Kanemura:2015ska,Draper:2020tyq,Gori:2017qwg}, 
electroweak phase transitions in the early universe
\cite{Basler:2016obg,Basler:2017uxn,Basler:2019iuu,Kainulainen:2019kyp,Basler:2021kgq},
precision calculations of Higgs decays
\cite{Krause:2016oke,Krause:2016xku,Altenkamp:2017kxk,Altenkamp:2017ldc,Altenkamp:2018hrq,Denner:2018opp,Kanemura:2018yai,Krause:2019qwe,Aiko:2021can}.
Remarkably, the \THDM\ is one of the very few single field extensions of the
Standard Model that can explain the deviation between
$\amu^\text{Exp}$ and $\amu^{\SM}$ \cite{Athron:2021iuf} while
satisfying existing constraints from collider physics searches and
other observables.

A second Higgs doublet that can couple to \SM\ fermions generically induces large tree-level flavour changing neutral currents at odds with experiment.  
Due to this \THDM s are often classified according to the discrete symmetries imposed to conserve flavour, giving four types \cite{Barger:1989fj}: type I \cite{Haber:1978jt}, type II \cite{Hall:1981bc}, type X (often called the lepton specific \THDM) \cite{Barnett:1983mm,Barnett:1984zy} and type Y (sometimes refereed to as the flipped \THDM) \cite{Barger:2009me,Logan:2010ag}.  
Large flavour changing neutral currents can also be avoided by assuming an alignment
in flavour space between the Yukawa matrices of the two doublets, giving the so-called
flavour-aligned \THDM\ (\FATHDM) \cite{Pich:2009sp,Tuzon:2010vt}.
After the first results of the LHC, Ref.\ \cite{Broggio:2014mna}
systematically investigated the phenomenology of all \THDM\ versions
with discrete symmetries and showed that among them, only the type X
variant is able to provide significant contributions to $\amu$. The
type X explanation  of the deviation between $\amu^\text{Exp}$ and $\amu^{\SM}$
has been further explored in Refs.\
\cite{Wang:2014sda,Abe:2015oca,Chun:2015xfx,Chun:2015hsa,Chun:2016hzs,Wang:2018hnw,Chun:2019oix,Chun:2019sjo,Keung:2021rps,Ferreira:2021gke,Han:2021gfu,Eung:2021bef,Jueid:2021avn,Dey:2021pyn}.
The more general flavour-aligned \THDM\ and its contributions to
$\amu$ were studied in
Refs.\ \cite{Ilisie:2015tra,Han:2015yys,Cherchiglia:2016eui,Cherchiglia:2017uwv,Li:2020dbg,Athron:2021iuf}. Other variants, including more general 
realisations that allow for Higgs-mediated flavour violation were investigated in Refs.\
\cite{Crivellin:2015hha,Omura:2015nja,Iguro:2019sly,Jana:2020pxx,Hou:2021sfl,Hou:2021qmf,Hou:2021wjj,Atkinson:2021eox}.

Phenomenological investigations of \BSM\ physics are greatly enhanced
by the use of precise software tools.  These tools can automate the
calculation of precision corrections, or provide numerical methods to
quickly and reliably solve related problems.  For example in the
\THDM\ one is often concerned that a particular benchmark point
respects measured limits of electroweak oblique parameters, or that
the new couplings and bosons do not provide contributions to Higgs
decays which violate collider constraints.  
For the \THDM\ a widely used software package is
\THDMC\ \cite{Eriksson:2009ws}, which provides calculations of the
spectrum, decays, oblique $S$, $T$ and $U$ parameters as well as a
calculation of the \THDM\ new physics contributions to $\amu$.  There
also exists the tool \ScannerS\ \cite{Muhlleitner:2020wwk} which can
place theoretical, experimental, dark matter, and electroweak phase
transition constraints on extended scalar sectors, including the
\THDM.  A more precise calculation of the decays is available from the
\THDM\ decay dedicated code \THDECAY\ \cite{Krause:2018wmo}, while
\PROPHECYfourF\ \cite{Denner:2019fcr} provides a package which focuses
on calculating $h\rightarrow WW/ZZ\rightarrow4f$ decays.
Additionally, the tools
\HiggsBounds\ \cite{Bechtle:2008jh,Bechtle:2011sb,Bechtle:2012lvg,Bechtle:2013wla,Bechtle:2015pma}
and \HiggsSignals\ \cite{Stal:2013hwa,Bechtle:2013xfa,Bechtle:2014ewa}
allow one to place constraints on the Higgs sectors of \BSM\ physics
through the measured behaviour of Higgs bosons from collider search
experiments.

Higher precision calculations for the spectrum, decays and other
observables in the \THDM\ are also available for arbitrary
user-defined extensions of the \SM\ through codes such as
\SARAH/\spheno\ \cite{Staub:2009bi,Staub:2010jh,Staub:2012pb,Staub:2013tta,Porod:2003um,Porod:2011nf}
and
\FS\ \cite{Athron:2014yba,Athron:2016fuq,Athron:2017fvs,Athron:2021kve}
(with decays recently added in \FD\ \cite{Athron:2021kve}) where model
files for the \THDM\ are already distributed.  Additionally, both
packages provide one-loop contributions to $\amuBSM$.  For two-loop
contributions to $\amuBSM$ in the \MSSM, \FS\ links to the dedicated
tool \GMTCalc\ \cite{Athron:2015rva}. For the \THDM\ there was no such
option until now.  Here we extend \GMTCalc\ with the \THDM\ to provide
a program which includes state of the art two-loop level contributions
of Ref.\ \cite{Cherchiglia:2016eui} to the anomalous magnetic moment
of the muon.

The \THDM\ contributions implemented in \GMTCalc\ version 2 include
the one-loop contributions, the two-loop fermionic corrections
(including the well-known Barr-Zee diagrams), and the complete set of
bosonic two-loop corrections. The one-loop contributions are of the
order $\order(m_\mu^4)$ and therefore usually subdominant. The
two-loop contributions arise  at
$\order(m_\mu^2)$ and are implemented at this order. 
Ref.\ \cite{Cherchiglia:2016eui} has assumed the flavour-aligned
\THDM\ with vanishing couplings $\lambda_{6,7}$, but here we relax
this assumption and allow the more general case.
The \THDM\ version of \GMTCalc~2 allows the user to input
deviations from the aligned limit, as well as select any one of the
\THDM\ types I, II, X or Y.  Just like version 1, \GMTCalc~2
allows the user to input parameter information using an SLHA-like
\cite{Skands:2003cj,Allanach:2008qq} input file.  We also
provide interfaces in C, C++, \python\ and \mathematica\ to make it easy
to link to other public codes.
Furthermore, \GMTCalc~2 (hereafter \GMTCalc) can be used as a standalone tool for studies
of $\amu$ in the \THDM, or to explore the \THDM\ phenomenology more
broadly it can be used in combination with other codes via the SLHA
interface.  For example, \GMTCalc\ can be called alongside \FS,
or in combination with other standalone tools like \THDECAY\ or
alongside the \THDMC\ package, replacing its native calculation of the
anomalous magnetic moment of the muon.  The \MSSM\ calculation is
already available in
\GAMBIT\ \cite{GAMBIT:2017yxo,GAMBITModelsWorkgroup:2017ilg} and it
should be straightforward to extend this to also use the \THDM\
calculation in future versions.

The rest of this paper is divided into the following sections.
\secref{sec:model} provides a pedagogical introduction to the \THDM, giving
the Lagrangian both before and after electroweak symmetry breaking,
and defining the Yukawa couplings in the various types of the \THDM.
There we give the one- and two-loop contributions to $\amuBSM$, and
the uncertainty estimate for the calculation.  
\secref{sec:implementation} describes various ways to use the \THDM\
in \GMTCalc.  It also shows how to interface \GMTCalc\ using C++,
C, \python\ and \mathematica.  Finally, \secref{sec:applications}
demonstrates various possible ways \GMTCalc\ can be applied to
calculate the \BSM\ contributions to the anomalous magnetic moment of
the muon in the \THDM.

\section{Review of 2HDM and implemented contributions to $\amu$} 
\label{sec:model}

\subsection{The Model}

The Two-Higgs Doublet Model (\THDM) extends the Standard Model (\SM)
with an additional scalar $\SU(2)$ doublet. We denote the two complex
Higgs $\SU(2)$ doublets in the \THDM\ as $\Phi_i$ ($i=1,2$),
\begin{align}
  \Phi_i =
  \begin{pmatrix}
    a_i ^+ \\ \frac{1}{\sqrt{2}} \left( v_i + b_i + \imath c_i \right)
  \end{pmatrix},
  \label{eq:higgsdoublets}
\end{align}
where $b_i$ and $c_i$ are real scalar fields and $a_i^+$ are complex scalar fields. 
Each Higgs doublet acquires a real non-zero vacuum expectation value (VEV), $v_i$, which satisfy
\begin{align}
  \tan\beta &= \frac{v_2}{v_1}, &
  v^2 &= v_1^2 + v_2^2,
  \label{eq:vevs} 
\end{align}
with $0 \leq \beta \leq \pi/2$ and $v \approx 246\GeV$, which implies
$v_1 = v\cos\beta$ and $v_2 = v\sin\beta$.
The extended Lagrangian includes the Higgs potential $\mathcal{L}_{\text{Scalar}}$ and the Yukawa interaction part $\mathcal{L}_{\text{Yuk}}$,
\begin{align}
  \mathcal{L} &\ni \mathcal{L}_\text{Scalar} + \mathcal{L}_\text{Yuk} . 
\end{align}
The most general form of Higgs potential is 
\begin{align}
\begin{split}
  -\mathcal{L}_\text{Scalar} ={}&
  m_{11}^2 \Phi_1^\dagger \Phi_1
  + m_{22}^2 \Phi_2^\dagger \Phi_2
  - \left[ m_{12}^2 \Phi_1^\dagger \Phi_2 + \hc \right] \\
  &+ \frac{1}{2} \lambda_1 \left( \Phi_1^\dagger \Phi_1 \right)^2
   + \frac{1}{2} \lambda_2 \left( \Phi_2^\dagger \Phi_2 \right)^2
   + \lambda_3 \left( \Phi_1^\dagger \Phi_1 \right)\left( \Phi_2^\dagger \Phi_2 \right)
   + \lambda_4 \left( \Phi_1^\dagger \Phi_2 \right)\left( \Phi_2^\dagger \Phi_1 \right) \\
  &+ \left[
    \frac{1}{2} \lambda_5 \left( \Phi_1^\dagger \Phi_2 \right)^2
    + \lambda_6 \left( \Phi_1^\dagger \Phi_1 \right)\left( \Phi_1^\dagger \Phi_2 \right)
    + \lambda_7 \left( \Phi_2^\dagger \Phi_2 \right)\left( \Phi_1^\dagger \Phi_2 \right)
    + \hc
  \right] , 
\end{split} 
\label{eq:Lscal}
\end{align}
where $\lambda_5$, $\lambda_6$, $\lambda_7$ and $m_{12} ^2$ are real for the $CP$-conserving Higgs potential.
The mass eigenstates of Higgs and Goldstone bosons $h$, $H$, $A$, $H^+$, $G^0$ and $G^+$ are obtained as
\begin{align}
  \begin{pmatrix} h \\ H \end{pmatrix} &= Z_{\mathcal{H}^0} \mathcal{H}^0 ,&
  \begin{pmatrix} G^0 \\ A \end{pmatrix} &= Z_{\mathcal{A}} \mathcal{A} ,&
  \begin{pmatrix} G^+ \\ H^+ \end{pmatrix} &= Z_{\mathcal{H}^+} \mathcal{H}^+ ,
  \label{eq:eigenstate1}
\end{align}
with 
\begin{align}
  \mathcal{H}^0 &=
  \begin{pmatrix}
    b_1 \\ b_2
  \end{pmatrix}, &
  \mathcal{A} &=
  \begin{pmatrix}
    c_1 \\ c_2
  \end{pmatrix} ,&
  \mathcal{H}^+ &=
  \begin{pmatrix}
    a_1^+ \\ a_2^+
  \end{pmatrix},
  \label{eq:gaugestate1}
\end{align}
from Eq.~\eqref{eq:Lscal}.  The orthogonal matrices
$Z_{\mathcal{H}^0}$,
$Z_{\mathcal{A}}$ and
$Z_{\mathcal{H}^\pm}$ for
Eq.~\eqref{eq:eigenstate1} are given as
\begin{align}
  Z_{\mathcal{H}^0} &=
  \begin{pmatrix}
    -\sin\alpha & \cos\alpha \\
    \cos\alpha & \sin\alpha
  \end{pmatrix} ,&
  Z_{\mathcal{A}} &=
  \begin{pmatrix}
    \cos\beta & \sin\beta \\
    -\sin\beta & \cos\beta
  \end{pmatrix} ,&
  Z_{\mathcal{H}^+} &=
  \begin{pmatrix}
    \cos\beta & \sin\beta \\
    -\sin\beta & \cos\beta
  \end{pmatrix} .
\end{align}
The corresponding mass square matrices $M_{\mathcal{H}^0} ^2$,
$M_{\mathcal{A}} ^2$ and $M_{\mathcal{H}^+} ^2$ for $\mathcal{H}^0$,
$\mathcal{A}$ and $\mathcal{H}^+$ in Eq.~\eqref{eq:gaugestate1},
respectively, are diagonalized as
\begin{align}
  M_{\mathcal{H}^0} ^2 &= Z_{\mathcal{H}^0}^\dagger (M_{\mathcal{H}^0}^D) ^2 Z_{\mathcal{H}^0} ,&
  M_{\mathcal{A}} ^2 &= Z_\mathcal{A}^\dagger (M_{\mathcal{A}}^D) ^2 Z_\mathcal{A} , &
  M_{\mathcal{H}^+} ^2 &= Z_{\mathcal{H}^+}^\dagger (M_{\mathcal{H}^+}^D) ^2 Z_{\mathcal{H}^+} .
\end{align}
The Higgs boson mass eigenvalues are denoted as
\begin{align}
  M_{\mathcal{H}^0}^D &= \diag( m_h , m_H ), &
  M_{\mathcal{A}}^D &= \diag( m_{G^0} , m_A ), &
  M_{\mathcal{H}^\pm}^D &= \diag( m_{G^\pm} , m_{H^\pm} ),
\end{align}
where in Feynman gauge $m_{G^+}=m_W$ and $m_{G^0}=m_Z$.
We chose the $CP$-even Higgs boson mixing angle $\alpha$ such that
$-\pi/2 \leq \beta - \alpha \leq \pi/2$ \cite{Eriksson:2009ws}.

The general form of the Yukawa interaction Lagrangian is given as
\begin{align}
  -\mathcal{L}_\text{Yuk} ={}&
    \Gamma_d ^0\overline{q_L ^0}\Phi_1 d_R ^0
  + \Gamma_u ^0 \overline{q_L ^0}\Phi_1 ^{c} u_R ^0
  + \Gamma_l ^0 \overline{l_L ^0}\Phi_1 e_R ^0
  + \PiL_d \overline{q_L ^0}\Phi_2 d_R ^0
  + \PiL_u \overline{q_L ^0}\Phi_2 ^{c} u_R ^0
  + \PiL_l \overline{l_L ^0}\Phi_2 e_R ^0
  + \hc,
  \label{eq:Lyuk}
\end{align}
where $\Phi_i^{c} \equiv \imath \sigma^2 \Phi_i^\ast$, and
$q_L ^0$, $l_L ^0$, $u_R ^0$, $d_R ^0$, and $e_R ^0$ are fermion gauge eigenstates.%
\footnote{The $\Gamma^0$ and $\PiL$ matrices correspond to $\eta_{1}$ and $\eta_{2}$ in Ref.\ \cite{Branco:2011iw}, see their Eq.~(92).} 
In general the Yukawa coupling matrices,  $\Gamma_f ^0$ and $\PiL_f$ ($f=u,d,l$)
are complex and non-diagonal $3\times 3$ matrices which
can not be simultaneously diagonalized, which produces flavour changing neutral currents (FCNC).

From the Yukawa interaction Lagrangian in Eq.~\eqref{eq:Lyuk} the $3\times 3$
fermion mass matrices $M_f$ ($f=u,d,l$) are obtained as 
\begin{align}
  \label{eq:massdefup}
  M_u &= \frac{1}{\sqrt{2}} \left( v_1 \Gamma_u ^0 + v_2 \PiL_u \right), \\
  \label{eq:massdefdown}
  M_d &= \frac{1}{\sqrt{2}} \left( v_1 \Gamma_d ^0 + v_2 \PiL_d \right), \\
  \label{eq:massdeflepton}
  M_l &= \frac{1}{\sqrt{2}} \left( v_1 \Gamma_l ^0+ v_2 \PiL_l \right).
\end{align}
The fermion mass matrices are diagonalized using singular value
decomposition. The diagonal fermion mass matrices $M_f^D$ ($f=u,d,l$)
are given by
\begin{align}
  \label{eq:massdiagup}
  M_u &= V_u^\dagger M_u^D U_u , & &\text{ with } & M_u^D &= \diag( m_u , m_c , m_t) ,& \\
  \label{eq:massdiagdown}
  M_d &= V_d^\dagger M_d^D U_d , & &\text{ with } & M_d^D &= \diag( m_d , m_s , m_b) ,& \\
  \label{eq:massdiaglepton}
  M_l &= V_l^\dagger M_l^D U_l , & &\text{ with } & M_l^D &= \diag( m_e , m_\mu , m_\tau) , 
\end{align}
where $V_f$ and $U_f$ are unitary matrices,
and the corresponding fermion mass eigenstates $f=u,d,l$ are obtained as 
\begin{align}
  f _L &= V _f f_L ^0, &\text{and}& &f _R =U _f f_R ^0 .&
  \label{eq:fmasseigen}
\end{align}
By combining Eq.~\eqref{eq:vevs} and Eqs.\
\eqref{eq:massdefup}--\eqref{eq:massdeflepton}, $\Gamma_u ^0$, $\Gamma_d ^0$
and $\Gamma_l ^0$ can be expressed in terms of $M_f$ and $\PiL _f$ as
\begin{equation} \label{eq:gthdmfixyukawas}
	\Gamma_f ^0 = \frac{\sqrt{2} M_f}{v\cos\beta} - \PiL_f \tan\beta.
\end{equation}

The Yukawa interaction Lagrangian Eq.~\eqref{eq:Lyuk} can be also
written in the so-called Higgs basis, where only one Higgs doublet has
non-zero VEV. This can be obtained by rotating $\Phi_{1,2}$ and the
Yukawa matrices as
\begin{align}
  \begin{pmatrix}
    \Phi _v \\
    \Phi _\perp
  \end{pmatrix}
  &=
  \begin{pmatrix}
    \cos\beta & \sin\beta \\
    -\sin\beta & \cos\beta 
  \end{pmatrix}
  \begin{pmatrix}
    \Phi _1 \\
    \Phi _2
  \end{pmatrix},
  \label{eq:Higgsbasis}
  \\
  \begin{pmatrix}
    \Sigma _f ^0\\
    \rhoL _f
  \end{pmatrix}
  &=
  \begin{pmatrix}
    \cos\beta & \sin\beta \\
    -\sin\beta & \cos\beta
  \end{pmatrix}
  \begin{pmatrix}
    \Gamma _f ^0 \\
    \PiL _f
  \end{pmatrix} ,
\label{eq:yukrelation}
\end{align}
such that only $\Phi_v$ has the VEV $v$ and $\Sigma^0_f$ are its Yukawa
couplings.  It contains the \SM-like Goldstone bosons, whereas the second field 
$\Phi_\perp$ contains non-\SM\ Higgs bosons $H^\pm$ and $A$, and its
Yukawa couplings $\rhoL_f$ are important parameters in the
\THDM.\footnote{The $\Sigma^0$ and $\rhoL$ matrices correspond to
  $\eta$ and $\hat{\xi}$ in Ref.\ \cite{Branco:2011iw}, see their
  Eq.~(30).} 
With this transformation the Yukawa interaction Lagrangian becomes
\begin{align}
  -\mathcal{L}_\text{Yuk} ={}&
    \Sigma_d ^0 \overline{q_L ^0}\Phi_v d_R ^0
  + \Sigma_u ^0 \overline{q_L ^0}\Phi_v ^{c} u_R ^0
  + \Sigma_l ^0 \overline{l_L ^0}\Phi_v e_R ^0
  + \rhoL_d \overline{q_L ^0}\Phi_\perp d_R ^0
  + \rhoL_u \overline{q_L ^0}\Phi_\perp ^{c} u_R ^0
  + \rhoL_l \overline{l_L ^0}\Phi_\perp e_R ^0
  + \hc
  \label{eq:Lyukv}
\end{align}
In this way, the fermion masses are purely generated from the Higgs
doublet $\Phi_v$, and the fermion mass matrices in
Eqs.~\eqref{eq:massdefup}--\eqref{eq:massdeflepton} become  
\begin{align}
  M_u &= \frac{v}{\sqrt{2}}\Sigma_u ^0 , &M_d &= \frac{v}{\sqrt{2}}\Sigma_d ^0 , &M_l &= \frac{v}{\sqrt{2}}\Sigma _l ^0,
  \label{eq:fermionmassv}
\end{align}
which implies $V_f \Sigma_f^0 U_f^\dagger = \sqrt{2}M_f^D/v$.
In parallel to Eqs.~\eqref{eq:massdiagup}--\eqref{eq:massdiaglepton}
we may define Yukawa matrices in the basis of mass eigenstates as ($f=u,d,l$)
\begin{subequations}
  \begin{align}
  \Pi _f &\equiv V _f \PiL _f U _f ^\dagger, &
  \Sigma _f &\equiv V_f \Sigma _f ^0 U _f ^\dagger, \\
  \Gamma _f &\equiv V _f \Gamma _f^0 U _f ^\dagger, &
  \rho _f& \equiv V _f \rhoL _f U _f ^\dagger.
  \end{align}
\end{subequations}
Importantly, in the Higgs basis the $\Sigma _f ^0$ are diagonalized by the singular value decomposition
involving $V_f$ and $U_f$, but in general the $\rhoL _f$ are not. This
implies that Higgs-mediated FCNCs are induced by the coupling with the zero-VEV Higgs doublet $\Phi _\perp$. 

From Eqs.~\eqref{eq:massdiagup}--\eqref{eq:massdiaglepton},
\eqref{eq:yukrelation}, and \eqref{eq:fermionmassv} we obtain
\begin{align}
  \Pi _f &= \sqrt{2}\frac{M_f ^D}{v} \sin\beta + \rho _f \cos\beta ,
  \label{eq:Piprime}
\end{align}
where $\Pi _f \equiv V _f \PiL _f U _f ^\dagger$ and
$\rho _f \equiv V _f \rhoL _f U _f ^\dagger$.  The $\rho _f$ matrices
contain non-zero off-diagonal components, which reflects that in
general the two types of Yukawa coupling matrices $\Gamma _f ^0$ and
$\PiL _f$ can not be simultaneously diagonalized in the \THDM. The
$\rho _f$ term in Eq.~\eqref{eq:Piprime} can induce Higgs-mediated
flavour-changing neutral currents (FCNC) at the tree-level.

\GMTCalc\ implements several variants of the \THDM\ with or without Higgs-mediated
FCNC:
\begin{itemize}
  \item The four well-known types with $Z_2$ symmetry: type I, type
    II, type X, type Y,
  \item
    the flavour-aligned \THDM\ (\FATHDM)~\cite{Pich:2009sp,Tuzon:2010vt},
    which also avoids tree-level Higgs-mediated FCNC and
    contains the four previous types as special cases,
  \item
    the \THDM\ with general Yukawa structures.
\end{itemize}
We will now describe these variants of the \THDM.  The most common way
to avoid FCNC is to impose a $Z_2$ symmetry, which leads to four
different types of Yukawa interactions in which specific subsets of
Yukawa couplings vanish. In the type I model all quarks and charged
leptons couple to $\Phi_2$, so we set
\begin{align}
  \Gamma_u ^0 = \Gamma_d ^0 = \Gamma_l ^0= 0.
\end{align}
In the type II model all up-type quarks couple to $\Phi_2$, while the
down-type quarks and charged leptons couple to $\Phi_1$, so we set
\begin{align}
  \Gamma_u ^0= \PiL_d = \PiL_l = 0.
\end{align}
In the type X model all quarks couple to $\Phi_2$, while the charged leptons
couple to $\Phi_1$, so we set
\begin{align}
  \Gamma_u ^0 = \Gamma_d ^0 = \PiL_l = 0.
\end{align}
In the type Y model the up-type quarks and charged leptons couple to $\Phi_2$, while the down-type quarks 
couple to $\Phi_1$, so we set
\begin{align}
  \Gamma_u ^0 = \PiL_d = \Gamma_l ^0 = 0.
\end{align}
In all these cases, trivially $\Gamma _f ^0$ and $\PiL _f$ can be
diagonalized simultaneously, and Higgs-mediated FCNC is avoided. 

In the flavour-aligned Two-Higgs Doublet Model (\FATHDM)~\cite{Pich:2009sp,Tuzon:2010vt} it is assumed that the $\PiL_f$ matrices are proportional to the $\Gamma_f ^0$ matrices, and we set
\begin{align}
  \label{eq:AlignedYuk}
  \PiL_f &= \begin{cases}
  	\xi_u^{*} \Gamma_u ^0, & \text{if }f=u, \\
  	\xi_{d,l} \Gamma_{d,l} ^0, & \text{if }f=d,l, \\
  \end{cases}  &
  &\text{with} &
  \xi_f &= \frac{\zeta_f + \tan\beta}{1 - \zeta_f\tan\beta} , 
\end{align}
where $\zeta _f$ are the alignment parameters which are constrained by experimental results and used in phenomenological studies. 
Note that the aligned model contains the type I, II, X and Y models as
special cases when the alignment parameters $\zeta_f$ take the values
given in \tabref{tab:zeta}.

We can summarize the fundamental Yukawa coupling matrices $\rho _f$
($f=u,d,l$) for the different \THDM\ variants and in different
parametrizations as follows:
\begin{align} \label{eq:rho}
  \rho_f =
  \begin{cases}
    \frac{\sqrt{2} M_f^D}{v} \zeta^{(*)}_f &\text{for type I, II, Y,
      Y (using  \tabref{tab:zeta})}
    \\
    &\text{and for the exact \FATHDM,}\\
    \frac{\sqrt{2} M_f^D}{v} \zeta^{(*)}_f +\Delta_f & \text{for the general \THDM\ in \FATHDM\ parametrization,} \\
    \frac{\Pi_f}{\cos\beta} - \frac{\sqrt{2} M_f^D}{v}\tan\beta & \text{for the general \THDM\ in $\Pi$ parametrization.} \\
  \end{cases}
\end{align}
Here the $\zeta_f$ are the parameters introduced in
Eq.~\eqref{eq:AlignedYuk} and \tabref{tab:zeta}, and the $(*)$
notation indicates the conjugate is present for the case $f=u$ and not
present for $f=d,l$.
\GMTCalc\ offers two parametrizations for the general \THDM. The 
``\FATHDM\ parametrization'' starts from the \FATHDM\ and allows to directly modify
the $\rho_f$ by additional matrices $\Delta_f$, which represents the
deviation from the flavour-aligned (or type I, II, X, Y) limit. 
The second ``$\Pi$ parametrization'' starts from Eq.~\eqref{eq:Piprime} and allows
to directly specify the fundamental Yukawa matrices $\Pi_f$.

\begin{table}[tb]
  \centering
  \caption{Values of the alignment parameters $\zeta_f$ for different types of Two-Higgs Doublet Models.}
  \begin{tabular}{rrrrr}
    \toprule
    & Type I & Type II & Type X & Type Y \\
    \midrule
    $\zeta_u$ & $\cot\beta$ &  $\cot\beta$ &  $\cot\beta$ &  $\cot\beta$ \\
    $\zeta_d$ & $\cot\beta$ & $-\tan\beta$ &  $\cot\beta$ & $-\tan\beta$ \\
    $\zeta_l$ & $\cot\beta$ & $-\tan\beta$ & $-\tan\beta$ &  $\cot\beta$ \\
    \bottomrule
  \end{tabular}
  \label{tab:zeta}
\end{table}

After suitable unitary transformations, the Yukawa interaction in the 
Lagrangian of Eq.\ \eqref{eq:Lyuk} takes the following form:
\begin{align}
  \label{eq:Lyuktransformed}
  \begin{split}
    -\mathcal{L}_\text{Yuk,int} ={}&
    H ^+ \bigg[\bar{u} \left(y_d^{\MHpm}\PR + y_u^{\MHpm}\PL \right) d + \bar{\nu} y_l^{\MHpm} \PR l \bigg] \\
    &+ \sum_{f=u,d,l} \left( \sum_{S=h,H}\,S{\bar{f}} y_f ^{S} \PR f - \imath \,A{\bar{f}} y_f ^{A} \PR f \right)
    + \hc,
  \end{split}
\end{align}
with the CKM matrix $\VCKM = V_u V_d^{\dagger}$ and the fermion mass
eigenstates $f=f_L + f_R$, $(f=u,d,l)$, defined in
Eq.~\eqref{eq:fmasseigen}. The coupling matrices $y_f^S$ are defined
as \cite{Cherchiglia:2016eui}:
\begin{align}
\label{eq:HiggsfermionCPevenh}
y_f^h &= \frac{M_f^D}{v} \sin(\beta - \alpha) + \frac{\rho_f}{\sqrt{2}} \cos(\beta -\alpha), \\
\label{eq:HiggsfermionCPevenH}
y_f^H &= \frac{M_f^D}{v} \cos(\beta - \alpha) - \frac{\rho_f}{\sqrt{2}} \sin(\beta - \alpha), \\
\label{eq:HiggsfermionCPodd}
y_f^A &= \begin{cases}
	\frac{\rho_u}{\sqrt{2}} & \text{if }f=u, \\
	- \frac{\rho_{d,l}}{\sqrt{2}} & \text{if }f=d,l, \\
\end{cases} 
\end{align}
and the coupling between the charged Higgs and each of the fermions is:
\begin{align}
\label{eq:HiggsfermionCharged}
y_f^{\MHpm} &= \begin{cases}
	-\rho_u^\dagger \VCKM & \text{if }f=u,\\
	\VCKM \rho_d & \text{if }f=d,\\
	\rho_l & \text{if }f=l, \\
\end{cases}
\end{align}
where $\rho_f$ 
is given in Eq.~\eqref{eq:rho}.

\subsection{Implemented contributions to $\amu$}

Relevant \THDM\ contributions to $\amu$ arise at the one-loop and the
two-loop level. The one-loop contributions are suppressed by two
additional powers of the muon Yukawa coupling and thus of
$\order(m_\mu^4)$; they are typically subleading unless some of the
non-\SM-like Higgs bosons have very small mass around a few GeV. The
two-loop contributions can be classified into fermionic and bosonic
contributions. They arise at the  $\order(m_\mu^2)$ and are typically
dominant. An important subclass  of two-loop contributions are the
so-called Barr-Zee diagrams, which contain a $\gamma\gamma$--Higgs
one-loop subdiagram. These diagrams have been studied extensively
\cite{Barr:1990vd,Chang:2000ii,Cheung:2001hz,Wu:2001vq,Krawczyk:2002df} and fully
calculated in Ref.\ \cite{Ilisie:2015tra}. The full set of two-loop
bosonic and fermionic diagrams (at $\order(m_\mu^2)$) has been
computed in Ref.\ \cite{Cherchiglia:2016eui} (see also
Ref.\ \cite{Cherchiglia:2017uwv} for further phenomenological
discussions of the individual contributions).

\GMTCalc\ implements the full set of one-loop \BSM\ contributions $\amu^\ol$ of
$\order(m_\mu^4)$ and the two-loop \BSM\ contributions $\amu^\tl$ of
$\order(m_\mu^2)$ in the \THDM\ from Ref.\ \cite{Cherchiglia:2016eui}.  The full
\BSM\ contribution in the \THDM\ (i.e.\ the difference between the
\THDM\ and the \SM\ contributions) calculated by \GMTCalc\ is therefore
the sum
\begin{align}
  \amuBSM=\amu^\ol+\amu^\tl.
\end{align}
In the following subsections we provide the explicit analytic results
implemented in \GMTCalc.

\subsubsection{One-loop contributions}

The one-loop \BSM\ contributions to $\amu^\ol$ in the \THDM\ is of
$\order(m_\mu^4)$ and are given by
\begin{align}
\begin{split}
  \amu^\ol = \frac{1}{8\pi^2} \Bigg\{
     & \sum_{i=1}^3 \Bigg[
         \sum_{S=h,H,A} \frac{m_\mu^2}{m_S^2} A_S(i, m_l, m_S^2, y_l^S)
         + \frac{m_\mu^2}{m_{H^\pm}^2} A_{H^\pm}(i, 0, m_{H^\pm}^2, y_l^{H^\pm})
     \Bigg] \\
     & - \frac{m_\mu^2}{m_\hSM^2} A_h(2, m_l, m_\hSM^2, \id)
  \Bigg\},
\end{split}\label{eq:amu1L}%
\end{align}
with $m_l = (m_e, m_\mu, m_\tau)$, $m_\nu = (m_{\nu_e}, m_{\nu_\mu}, m_{\nu_\tau})$ and
\begin{align} 
  \label{eqn:onelooph}
  A_h(i,m_l,m_S^2,y_l^S) ={}& A_S^+(i,m_l,m_S^2,y_l^S), \\
  \label{eqn:oneloopH}
  A_H(i,m_l,m_S^2,y_l^S) ={}& A_S^+(i,m_l,m_S^2,y_l^S), \\
  \label{eqn:oneloopA}
  A_A(i,m_l,m_S^2,y_l^S) ={}& A_S^-(i,m_l,m_S^2,y_l^S), \\
  \nonumber
  A_S^\pm(i,m_l,m_S^2,y_l^S) ={}&
    \frac{1}{24} \left( \left|(y_l^S)_{i2}\right|^2 + \left|(y_l^S)_{2i}\right|^2 \right) F_1^C\left(\frac{(m_l)^2_i}{m_S^2}\right) \\
    \label{eqn:oneloopS}
    &\pm \frac{1}{3} \real\left[(y_l^S)_{i2}^* (y_l^S)_{2i}^*\right] \frac{(m_l)_i}{m_\mu} F_2^C\left(\frac{(m_l)^2_i}{m_S^2}\right), \\
  \label{eqn:oneloopHp}
  A_{H^\pm}(i, m_\nu, m_S^2, y_l^S) ={}&
    -\frac{1}{48} \left|(y_l^S)_{i2}\right|^2 \left[ F_1^N \left(\frac{m_{\nu_\mu}^2}{m_S^2}\right) + F_1^N\left(\frac{(m_\nu)_i^2}{m_S^2}\right) \right].
\end{align}
Note, that the one-loop contribution from the \SM\ Higgs boson $\hSM$
is subtracted in Eq.~\eqref{eq:amu1L} and is therefore not included in
$\amu^\ol$.  The loop functions $F_1^C$, $F_2^C$ and $F_1^N$ are given
by
\begin{align}
  F_1^C(x) &= \frac{2}{(x - 1)^4}\left(2 + 3x - 6x^2 + x^3 + 6x\ln x\right), & F_1^C(0)&=4, & F_1^C(1)&=1, \\
  F_2^C(x) &= \frac{3}{2(1 - x)^3}\left(-3 + 4x - x^2 - 2\ln x\right), & F_2^C(1)&=1, & F_2^C(\infty)&=0, \\
  F_1^N(x) &= \frac{2}{(x - 1)^4}\left(1 - 6x + 3x^2 + 2x^3 - 6x^2\ln x\right), & F_1^N(0)&=2, & F_1^N(1)&=1, & F_1^N(\infty)=0.
\end{align}

\subsubsection{Two-loop contributions}

The two-loop contributions of $\order(m_\mu^2)$ are divided into
fermionic and bosonic loop contributions according to Ref.\
\cite{Cherchiglia:2016eui} as
\begin{align}
  \amu^{\tl} = \amuB + \amuF ,
  \label{eq:amu2L}
\end{align}
where the very small contribution from the shift of the Fermi constant,
$\amu^{\Delta r\text{-shift}}$, is neglected.  We generalize the
fermionic two-loop contributions from Ref.\ \cite{Cherchiglia:2016eui}
in the general \THDM\ to the case of CKM mixing as follows:
\begin{align}
  \amuF &= \amu^{\text{F,N}} + \amu^{\text{F,C}} , \\
  \amu^{\text{F,N}} &= \sum_{f=u,d,l} \sum_{i=1}^3 \frac{\aem^2 m_\mu^2}{4\pi^2 m_W^2 s_W^2} \left[
  \sum_{S=h,H,A} f_f^S(m_S,(m_f)_i) \frac{\real\left[(y_f^S)_{ii}^* (y_l^S)_{22}\right] v^2}{(m_f)_i m_\mu}
  - f_f^{\hSM}(m_{\hSM},(m_f)_i) \right], \label{eq:amu2LFN} \\
  \amu^{\text{F,C}} &=  \sum_{i,j=1}^3
  \frac{\aem^2 m_\mu^2}{32\pi^2 m_W^2 s_W^4}
  \Bigg[
    f_u^{H^\pm}(m_{H^\pm},(m_u)_i,(m_d)_j)
    \frac{\real \left[(y_u^{H^\pm})_{ij}^* (\VCKM)_{ij}(y_l^{H^\pm})_{22}\right] v^2}{2 (m_u)_i
      m_\mu}
    \nonumber\\
    &\qquad\qquad\qquad\qquad\qquad
    +f_d^{H^\pm}(m_{H^\pm},(m_d)_j,(m_u)_i)
    \frac{\real \left[(y_d^{H^\pm})_{ij}^* (\VCKM)_{ij}(y_l^{H^\pm})_{22}\right] v^2}{2 (m_d)_j
      m_\mu}
    \nonumber\\
    &\qquad\qquad\qquad\qquad\qquad
    +f_l^{H^\pm}(m_{H^\pm},(m_l)_i,0)
    \frac{\real\left[ (y_l^{H^\pm})_{ij}^* \delta_{ij}(y_l^{H^\pm})_{22}\right] v^2}{2 (m_l)_i
      m_\mu}
    \Bigg].
\end{align}

Note, that the two-loop
contribution from the \SM\ Higgs boson $\hSM$ is subtracted in
Eq.~\eqref{eq:amu2LFN} and is therefore not included in $\amu^\tl$.
The loop functions $f_f^S$ ($S=h,H,A,\hSM$) are defined as:
\begin{align}
	\label{eqn:twoloopfS}
	f_f^S(m_S,m_f) ={}& q_f^2 N^c_f \frac{m_f^2}{m_S^2} \F_S (m_S,m_f)
	- q_f N^c_f \frac{g^l_v g^f_v}{s_W^2 c_W^2} \frac{m_f^2}{m_S^2- m_Z^2} \left[\F_S (m_S,m_f) - \F_S (m_Z,m_f)\right], \\
	\label{eqn:twoloopFS}
	\F_S(m_S,m_f) ={}& \begin{cases}
		-2 + \ln\bigg(\frac{m_S^2}{m_f^2}\bigg) - \bigg(\frac{m_S^2-2m_f^2}{m_S^2}\bigg) \frac{\Phi(m_S,m_f,m_f)}{m_S^2-4m_f^2}, & S=h,H,\hSM,\\
		\frac{\Phi(m_S,m_f,m_f)}{m_S^2-4m_f^2}, & S=A,\\
	\end{cases}
\end{align}
where $N^c_f = (1,3,3)$, $g^f_v = T^3_f/2 - s_W^2 q_f$,
$T^3_f = (-1/2,-1/2,1/2)$ and $q_f=(-1,-1/3,2/3)$ for $f=(l,d,u)$.
The loop function $f_f^{H^\pm}$ is defined as:
\begin{align}
	\label{eqn:twoloopHp}
	f_f^{H^\pm}(m_{H^\pm},m_f,m_{f'}) = \begin{cases}
		\frac{m_l^2}{m_{H^\pm}^2-m_W^2} \left[\F_l^{H\pm}\bigg(\frac{m_l^2}{m_{H\pm}^2}\bigg) - \F_l^{H^\pm}\bigg(\frac{m_l^2}{m_W^2}\bigg)\right], & f=l, f'=\nu, \\
		3\frac{m_d^2}{m_{H^\pm}^2-m_W^2} \left[\F_d^{H\pm}\bigg(\frac{m_d^2}{m_{H\pm}^2},\frac{m_u^2}{m_{H\pm}^2}\bigg) - \F_d^{H^\pm}\bigg(\frac{m_d^2}{m_W^2},\frac{m_u^2}{m_W^2}\bigg)\right], & f=d, f'=u, \\
		3\frac{m_u^2}{m_{H^\pm}^2-m_W^2} \left[\F_u^{H\pm}\bigg(\frac{m_d^2}{m_{H\pm}^2},\frac{m_u^2}{m_{H\pm}^2}\bigg) - \F_u^{H^\pm}\bigg(\frac{m_d^2}{m_W^2},\frac{m_u^2}{m_W^2}\bigg)\right], & f=u, f'=d,
	\end{cases}
\end{align}
where $f'_j$ is the $\SU(2)_L$ partner of generation $j$ of the
fermion $f_i$ from generation $i$, and
\begin{align}
	\label{eqn:twoloopFlHp}
	\F_l^{H^\pm}(x_l) ={}& x_l + x_l (x_l-1) \left[\dilog\bigg(1-\frac{1}{x_l}\bigg) -\frac{\pi^2}{6}\right] + \left(x_l-\frac{1}{2}\right)\ln(x_l), \\ 
	\nonumber
	\F_d^{H^\pm}(x_d,x_u) ={}& -(x_u+x_d) + \left[\frac{\overline{c}}{y} - \frac{c(x_u-x_d)}{y}\right] \Phi(x_d,x_u,1) \\ 
	\nonumber
	&+ c\left[\dilog(1-x_d/x_u) - \frac{1}{2} \ln{(x_u)} \ln{(x_d/x_u)}\right] \\
	\label{eqn:twoloopFdHp}
	&+ (s+x_d)\ln{(x_d)} + (s-x_u)\ln{(x_u)}, \\ 
	\nonumber
	\F_u^{H^\pm}(x_d,x_u) ={}& \F_d^{H^\pm}(x_d,x_u) (q_u \rightarrow 2+q_u, q_d \rightarrow 2+q_d) \\
	\label{eqn:twoloopFuHp}
	&-\frac{4}{3} \frac{(x_u-x_d-1)}{y} \Phi(x_d,x_u,1) - \frac{1}{3} \left[\ln^2{(x_d)} - \ln^2{(x_u)}\right],
\end{align}
and
\begin{align}
  c &= (x_u - x_d)^2 - q_u x_u + q_d x_d, \\
  \overline{c} &= (x_u - q_u) x_u - (x_d + q_d) x_d, \\
  y &= (x_u - x_d)^2 - 2 (x_u + x_d) + 1, \\
  s &= \frac{q_u+q_d}{4}.  
\end{align}
The loop function $\Phi(m_1^2,m_2^2,m_3^2)$ \cite{DAVYDYCHEV1993123}
is defined as:
\begin{align}
	\Phi(m_1^2,m_2^2,m_3^2) &= \frac{\lambda}{2} \bigg[2\ln{(\alpha_+)}\ln{(\alpha_-)} - \ln{\bigg(\frac{m_1^2}{m_3^2}\bigg)}\ln{\bigg(\frac{m_2^2}{m_3^2}\bigg)} - 2\dilog{(\alpha_+)} - 2\dilog{(\alpha_-)} + \frac{\pi^2}{3}\bigg], \\
	\alpha_{\pm} &= \frac{m_3^2 \pm m_1^2 \mp m_2^2 - \lambda}{2m_3^2}, \\
	\lambda &= \sqrt{m_1^4 + m_2^4 + m_3^4 - 2m_1^2 m_2^2 - 2m_2^2 m_3^2 -2 m_3^2 m_1^2}.  
\end{align}

The bosonic two-loop contributions to the general \THDM\ are composed of three parts as follows:
\begin{equation} \label{eqn:twoloopB}
	\amuB = \amu^{\text{B,EW add}} + \amu^{\text{B,Yuk}} + \amu^{\text{B,non-Yuk}}.
\end{equation}
The two terms $\amu^{\text{B,EW add}}$ and $\amu^{\text{B,nonYuk}}$
correspond to diagrams with the \SM-like Higgs boson and diagrams
without the new Yukawa couplings, respectively. They are typically
subdominant, and full details on their definition and phenomenological
impact are given in 
Refs.~\cite{Cherchiglia:2016eui} and \cite{Cherchiglia:2017uwv}. 
The Yukawa contribution is given by
\begin{align} \label{eqn:twoloopYuk}
	\nonumber
	\amu^{\text{B,Yuk}} ={}& \frac{\aem^2}{574\pi^2c_W^4s_W^4} \frac{m_\mu^2}{m_Z^2} \bigg\{ a_{0,0}^{0} + a_{0,z}^{0} \left(\tan\beta-\frac{1}{\tan\beta}\right) \zeta_l + a_{5,0}^{0} \Lambda_5 + a_{5,z}^{0} \left(\tan\beta-\frac{1}{\tan\beta}\right) \Lambda_{567} \zeta_l \\
	&+ \left[a_{0,0}^{1} \left(\tan\beta-\frac{1}{\tan\beta}\right) + a_{0,z}^{1} \zeta_l + a_{5,0}^{1} \left(\tan\beta-\frac{1}{\tan\beta}\right) \Lambda_{567} + a_{5,z}^{1} \Lambda_5 \zeta_l\right] \cos(\beta - \alpha)\bigg\}, \\
	\Lambda_5 ={}& \frac{2m_{12}^2}{v^2\sin\beta\cos\beta}, \\
\label{eq:lambda567}        \Lambda_{567}={}& \Lambda_5+\frac{1}{\tan\beta-\frac{1}{\tan\beta}}\left(\frac{\lambda_6}{\sin^2\beta}-\frac{\lambda_7}{\cos^2\beta}\right),
\end{align}	
where the equations for $a_{\lambda,t}^{\eta}$ (with the common
prefactor put in the front of the above Eq.~\eqref{eqn:twoloopYuk})
are given in the appendix of
Ref.~\cite{Cherchiglia:2016eui}. Compared to this reference, the
expression has been generalized to include $\lambda_6$ and
$\lambda_7$.\footnote{%
  These two Higgs potential parameters only enter via triple Higgs
  couplings, and they enter only via the combination $\Lambda_{567}$
  defined in Eq.\ \eqref{eq:lambda567}. The suitable generalizations of Eqs.\ (3.25)--(3.26) of
  Ref.\ \cite{Cherchiglia:2016eui} for the required triple Higgs
  couplings can be obtained e.g.\ from formulas in the Appendix of
  Ref.\ \cite{Gunion:2002zf} and can simply be written as
    \begin{align}
g_{\Mh,H^\pm,H^\mp} \propto& \left\{\vac\,\left( \Lambda_5 - \frac{\MMh^2}{\vac^2} - 2\frac{\MMHpm^2}{\vac^2}\right)
+\, \eta\,\left(\tan\beta - \frac{1}{\tan\beta}\right)\frac{\vac}{2}\left(2\,\frac{\MMh^2}{\vac^2} - \Lambda_{567}\right)\right\},\\
g_{\MH,H^\pm,H^\mp} \propto& \left\{
\left(\tan\beta - \frac{1}{\tan\beta }\right)\frac{\vac}{2}\left( \Lambda_{567} -  2\,\frac{\MMH^2}{\vac^2} \right)
+ \eta \vac\,\left( \Lambda_5 - \frac{\MMH^2}{\vac^2} - 2\frac{\MMHpm^2}{\vac^2}\right)\right\}.
    \end{align}
    This structure clarifies the appearance of $\Lambda_{567}$ in
    Eq.\ \eqref{eqn:twoloopYuk}.
}
These bosonic contributions are implemented in the realistic
approximation of small $\cos(\beta-\alpha)$, and only terms up to
linear order in $\cos(\beta-\alpha)$ are taken into
account.  In this scenario the boson $h$ has the tree-level couplings of the \SM\ Higgs boson \cite{Gunion:2002zf}.  
Furthermore, the bosonic contributions are derived only for
the flavour-aligned \THDM. Referring to the different cases of Yukawa
couplings in Eq.\ \eqref{eq:rho}, the bosonic corrections apply for
the cases of the discrete symmetries, for the \FATHDM, and also for
the general \THDM\ in \FATHDM\ parametrization (assuming the $\Delta_f$
matrices are small). In contrast,  we set
$\zeta_l=0$ in Eq.~\eqref{eqn:twoloopYuk} in the case of the general
\THDM\ in $\Pi$ parametrization since in this case the bosonic
corrections are not applicable in this form.

The bosonic two-loop contributions
are also available in the \GMTCalc\ source code and in the form of a
\mathematica\ file.  

\subsubsection{Running couplings}
\label{sec:running_couplings}

Among the fermionic two-loop contributions, the diagrams with an
internal top quark, bottom quark or tau lepton loop give the dominant
contribution to $\amuBSM$.  These diagrams are proportional to the values
of the fermion masses in the loop.  So long as there is no three-loop
calculation available in the \THDM, it is formally irrelevant which
renormalization scheme to use for the fermion masses in the loop.  In \GMTCalc\ two possible
definitions of these fermion masses are implemented:
\begin{align}
  \text{Input masses:}&   \qquad m_t, m_b^\MS(m_b^\MS), m_\tau
  \label{eq:input_masses} \\
  \text{Running masses:}& \qquad m_t^\MS(Q), m_b^\MS(Q), m_\tau^\MS(Q)
  \label{eq:running_masses}
\end{align}
In the ``input masses'' scheme, the top quark pole mass $m_t$, the
$\MS$ bottom quark mass in the \SM\ with five active quark flavours,
$m_b^\MS$, at the renormalization scale $Q=m_b^\MS$, and the tau
lepton pole mass $m_\tau$ are used in the loops.  These masses are
typically used as input for spectrum generators
\cite{Skands:2003cj,Allanach:2008qq}.
In the ``running masses'' scheme, the running $\MS$ top quark mass
$m_t^\MS(Q)$, the $\MS$ bottom quark mass $m_b^\MS(Q)$ and the $\MS$
tau lepton mass $m_\tau^\MS(Q)$ are used in the loops.  The
renormalization scale $Q$ is set to the mass of the Higgs boson in the
Feynman diagram.  The ``running masses'' scheme is also used in
\THDMC\ \cite{Eriksson:2009ws}.  The difference between these two
schemes is shown in \secref{sec:applications}.

\subsubsection{Uncertainty estimate}

We provide an uncertainty estimate for $\amu^{\tl}$ as follows:
\begin{align} 
  \Damu^{\tl} &= \Damu^{\tl,\Delta r} + \Damu^{\tl,m_\mu^4} + \Damu^{\thl}, \label{eq:damu}
\end{align}
where
\begin{align}
  \Damu^{\tl,\Delta r} &= 2\times 10^{-12}, \\
  \Damu^{\tl,m_\mu^4} &= \left| \amu^{\ol}\;\Delta\aem \right|, \\
  \Damu^{\thl} &= \left| \amu^{\tl}\;\Delta\aem \right|,
\end{align}
and
\begin{align}
  \Delta\aem &= -\frac{4\aem}{\pi} \ln\left(\frac{m_\text{NP}}{m_\mu}\right),
  & m_\text{NP} &= \min\{m_H, m_A, m_{H^\pm}\}.
\end{align}
The term $\Damu^{\tl,\Delta r}$ accounts for the fact that the
two-loop contribution $\amu^{\Delta r\text{-shift}}$ has been
neglected in Eq.~\eqref{eq:amu2L}.  In Refs.~\cite{Lopez-Val:2012uou,Cherchiglia:2016eui} it was
shown that in the relevant parameter space
$|\amu^{\Delta r\text{-shift}}| \leq 2\times 10^{-12}$, which we use as
an upper bound in the uncertainty estimate.
The term $\Damu^{\tl,m_\mu^4}$ estimates missing two-loop terms of
$\order(m_\mu^4)$ using the known universal two-loop QED logarithmic
contributions \cite{Degrassi:1998es}.
The term $\amu^{\thl}$ is an estimate for the expected three-loop
contributions.  From experience the QED contributions are among the
largest contributions at each loop level. For this reason we use again the known universal
logarithmic QED contributions from Ref.~\cite{Degrassi:1998es} to 
provide an estimate for the unknown three-loop contributions.

\section{Program details and usage} 
\label{sec:implementation}

\subsection{Quick start}

\GMTCalc\ can be downloaded from \url{https://gm2calc.hepforge.org/}
or \url{https://github.com/GM2Calc/GM2Calc}, for example as%
\footnote{To track changes made to \GMTCalc\ and get automatic
  updates one may alternatively clone the \lstinline|git| repository
  \url{https://github.com/GM2Calc/GM2Calc}}
\begin{lstlisting}[language=bash]
wget --content-disposition https://github.com/GM2Calc/GM2Calc/archive/v2.0.0.tar.gz
tar -xf GM2Calc-2.0.0.tar.gz
cd GM2Calc-2.0.0
\end{lstlisting}
To build \GMTCalc\ run the following commands:
\begin{lstlisting}[language=bash]
mkdir build
cd build
cmake ..
make
\end{lstlisting}
To calculate $\amuBSM$ in the \THDM\ with \GMTCalc\ from the command line,
run
\begin{lstlisting}[language=bash]
bin/gm2calc.x --thdm-input-file=../input/example.thdm
\end{lstlisting}
Here, \lstinline|example.thdm| is the name of the file that contains
the input parameters.\footnote{For instructions and examples of
  running the \MSSM\ evaluation see Ref.\ \cite{Athron:2015rva}.}

\subsection{Requirements}

To build \GMTCalc\ the following programs and libraries are required:
\begin{itemize}
\item C++14 and C11 compatible compilers
\item Eigen library, version 3.1 or higher \cite{eigenweb}
  [\url{http://eigen.tuxfamily.org}]
\item Boost library, version 1.37.0 or higher \cite{BoostLibrary}
  [\url{http://www.boost.org}]
\item (optional) \lstinline|Wolfram| \mathematica\ or \lstinline|Wolfram Engine|
  \cite{Mathematica} 
\item (optional) \python\ 2 or \python\ 3 \cite{Python}, using the package
  \lstinline|cppyy| \cite{cppyy}
  [\url{https://pypi.org/project/cppyy/}] 
\end{itemize}

\subsection{Running \GMTCalc\ from the command line}

\GMTCalc\ can be run from the command line with an SLHA-like
\cite{Skands:2003cj,Allanach:2008qq} input file as
\begin{lstlisting}[language=bash]
bin/gm2calc.x --thdm-input-file=example.thdm
\end{lstlisting}
where \lstinline|example.thdm| is the name of the SLHA-like input file.
Alternatively, the input parameters can be piped in an SLHA-like
format into \GMTCalc\ as
\begin{lstlisting}[language=bash]
cat example.thdm | bin/gm2calc.x --thdm-input-file=-
\end{lstlisting}
The calculated value for $\amuBSM$ and $\Damu$ is written to the standard
output.  Depending on the input options and parameters, the output may
contain the following block with $\amuBSM$ and $\Damu$:
\begin{lstlisting}[language=bash]
Block GM2CalcOutput
     0     1.67323025E-11   # Delta(g-2)_muon/2
     1     3.36159655E-12   # uncertainty of Delta(g-2)_muon/2
\end{lstlisting}
The SLHA-like format for the input parameters is defined in the
following subsections.

\subsubsection{General options}

In the SLHA-like input the selection of the configuration flags for
\GMTCalc\ can be given in the \lstinline|GM2CalcConfig| block as
defined in Ref.\ \cite{Athron:2015rva}.  An example
\lstinline|GM2CalcConfig| block reads as follows:
\begin{lstlisting}
Block GM2CalcConfig
     0     4     # output format (0 = minimal, 1 = detailed,
                 #  2 = NMSSMTools, 3 = SPheno, 4 = GM2Calc)
     1     2     # loop order (0, 1 or 2)
     2     1     # disable/enable tan(beta) resummation (0 or 1)
     3     0     # force output (0 or 1)
     4     0     # verbose output (0 or 1)
     5     1     # calculate uncertainty (0 or 1)
     6     1     # running couplings in the THDM
\end{lstlisting}
The entry \lstinline|GM2CalcConfig[0]| defines the output format.  If
\lstinline|GM2CalcConfig[0] = 0|, a single number is printed to the
\lstinline|stdout|.  This number is the value of $\amuBSM$ or the
uncertainty $\Damu$, depending on the value of
\lstinline|GM2CalcConfig[5]|: If \lstinline|GM2CalcConfig[5] = 0|, the
value of $\amuBSM$ is printed.  If \lstinline|GM2CalcConfig[5] = 1|, the
value of $\Damu$ is printed.  If \lstinline|GM2CalcConfig[0] = 1|,
a detailed output, suitable for debugging, is printed.  If
\lstinline|GM2CalcConfig[0] = 2|, the value of $\amuBSM$ is written to the
output block entry \lstinline|LOWEN[6]|.  If
\lstinline|GM2CalcConfig[0] = 3|, the value of $\amuBSM$ is written to the
output block entry \lstinline|SPhenoLowEnergy[21]|.  If
\lstinline|GM2CalcConfig[0] = 4| (default), the value of $\amuBSM$ is
written to the output block entry \lstinline|GM2CalcOutput[0]|.

The entry \lstinline|GM2CalcConfig[1]| defines the loop order of the
calculation, which can be set to \lstinline|0|, \lstinline|1| or
\lstinline|2|.  The default value is \lstinline|2| (recommended),
which corresponds to the two-loop calculation of $\amuBSM$.

The entry \lstinline|GM2CalcConfig[2]| disables/enables the
resummation of $\tan\beta$ in the \MSSM.  By default $\tan\beta$
resummation is enabled, which corresponds to
\lstinline|GM2CalcConfig[2] = 1|.  To disable $\tan\beta$ resummation,
set \lstinline|GM2CalcConfig[2] = 0|.  When the calculation is
performed in the \THDM, the value of \lstinline|GM2CalcConfig[2]| is
ignored.

The next two options are useful for debugging.  
By setting the entry \lstinline|GM2CalcConfig[3] = 1| (default:
\lstinline|0|), the output of \GMTCalc\ can be forced, even if a
physical problem (e.g.\ a tachyon) has occurred.  Warning: If a
physical problem has occurred, the output cannot be trusted.  Forcing
the output should only be used for debugging.
By setting the entry \lstinline|GM2CalcConfig[4] = 1| (default:
\lstinline|0|), additional information about the model parameters and
the calculation is printed.  

With the entry \lstinline|GM2CalcConfig[5]| the calculation of the
uncertainty $\Damu$, c.f.\ Eq.~\eqref{eq:damu}, can be
disabled/enabled (\lstinline|0| or \lstinline|1|).

The entry \lstinline|GM2CalcConfig[6]| (default: \lstinline|1|) is new
in \GMTCalc\ 2.0.0 and controls the definition of the fermion masses
that are inserted into the fermionic two-loop contribution $\amuF$,
see \secref{sec:running_couplings}.  If \lstinline|GM2CalcConfig[6] = 0|,
the input masses \eqref{eq:input_masses} are used.  If
\lstinline|GM2CalcConfig[6] = 1|, the running masses
\eqref{eq:running_masses} are used.  The difference between these two
schemes is shown in \secref{sec:applications}.

\subsubsection{Standard Model input parameters}

The Standard Model input parameters are read from the
\lstinline|SMINPUTS|, \lstinline|GM2CalcInput| and \lstinline|VCKMIN|
blocks.  Example blocks that define the Standard Model input
parameters may read:
\begin{lstlisting}
Block SMINPUTS
     1     128.94579        # alpha_em(MZ)^(-1) SM MS-bar
     3     0.1184           # alpha_s(MZ) SM MS-bar
     4     91.1876          # MZ(pole)
     5     4.18             # mb(mb) SM MS-bar
     6     173.34           # mtop(pole)
     7     1.77684          # mtau(pole)
     8     0                # mnu3(pole)
     9     80.385           # mW(pole)
    11     0.000510998928   # melectron(pole)
    12     0                # mnu1(pole)
    13     0.1056583715     # mmuon(pole)
    14     0                # mnu2(pole)
    21     0.0047           # md(2 GeV)
    22     0.0022           # mu(2 GeV)
    23     0.096            # ms(2 GeV)
    24     1.28             # mc(2 GeV)
Block GM2CalcInput
    33     125.09           # SM Higgs boson mass
Block VCKMIN                # CKM matrix in Wolfenstein parametrization
     1     0.2257           # lambda
     2     0.814            # A
     3     0.135            # rho-bar
     4     0.349            # eta-bar
\end{lstlisting}
The entries of the \lstinline|SMINPUTS| and \lstinline|VCKMIN| are
defined in Refs.\ \cite{Skands:2003cj,Allanach:2008qq}.  In the block
entry \lstinline|GM2CalcInput[33]| the mass of the Standard Model
Higgs boson can be given.  Unset parameters in the
\lstinline|SMINPUTS| and \lstinline|GM2CalcInput| blocks are
assigned default values.  Unset parameters in the \lstinline|VCKMIN|
block are assumed to be zero.

\subsubsection{Two-Higgs Doublet Model input parameters}

We allow the specification of the \THDM\ input parameters in two
different ``bases'', see \tabref{tab:basis}.  The basis parameters are
defined as in \cite{Eriksson:2009ws}.  In the ``gauge basis'' the
Lagrangian parameters, $\lambda_{1, \ldots, 7}$
are used as input.  In the ``mass basis'' the
Higgs boson masses and the mixing parameter $\sin(\beta-\alpha)$ are
used as input, instead of the Lagrangian parameters $\lambda_{1, \ldots, 5}$,
but $\lambda_6$ and $\lambda_7$ can still be used.
All available input parameters are listed in
\tabref{tab:basis2}.
\begin{table}[tb]
  \centering
  \caption{\THDM\ input parameters for different basis ($f=u,d,l$).}
  \begin{tabular}{ll}
    \toprule
    basis    & input parameters \\
    \midrule
    gauge  & $\lambda_1$, \ldots, $\lambda_7$, $\tan\beta$, $m_{12}^2$, $\zeta_f$, $\Delta_f$, $\Pi_f$ \\
    mass   & $m_h$, $m_H$, $m_A$, $m_{H^\pm}$, $\sin(\beta-\alpha)$, $\lambda_6$, $\lambda_7$, $\tan\beta$, $m_{12}^2$, $\zeta_f$, $\Delta_f$, $\Pi_f$ \\
    \bottomrule
  \end{tabular}
  \label{tab:basis}
\end{table}

\paragraph{Mass basis input parameters}

\begin{table}[tb]
  \centering
  \caption{\THDM\ input parameters for the SLHA and \mathematica\ interface.}
  \begin{tabular}{llll}
    \toprule
    input parameter & SLHA entry & \mathematica\ symbol & allowed values \\
    \midrule
    \multicolumn{4}{c}{parameters specific to the gauge basis} \\
    \midrule
    $\lambda_1,\ldots,\lambda_7$ & \lstinline|MINPAR[11]|, \ldots, \lstinline|MINPAR[17]| & \lstinline|lambda| & $\mathbb{R}$ \\
    \midrule
    \multicolumn{4}{c}{parameters specific to the mass basis} \\
    \midrule
    $\lambda_6$ & \lstinline|MINPAR[16]| & \lstinline|lambda6| & $\mathbb{R}$ \\
    $\lambda_7$ & \lstinline|MINPAR[17]| & \lstinline|lambda7| & $\mathbb{R}$ \\
    $\sin(\beta-\alpha)$ & \lstinline|MINPAR[20]| & \lstinline|sinBetaMinusAlpha| & $[-1,1]$ \\
    $\{m_h,m_H\}$ & \lstinline|MASS[25]|, \lstinline|MASS[35]| & \lstinline|Mhh| & $\{\mathbb{R}_{\geq 0},\mathbb{R}_{\geq 0}\}$ \\
    $m_A$ & \lstinline|MASS[36]| & \lstinline|MAh| & $\mathbb{R}_{\geq 0}$ \\
    $m_{H^\pm}$ & \lstinline|MASS[37]| & \lstinline|MHp| & $\mathbb{R}_{\geq 0}$ \\
    \midrule
    \multicolumn{4}{c}{parameters common to both the gauge and mass basis} \\
    \midrule
    Yukawa type & \lstinline|MINPAR[24]| & \lstinline|yukawaType| & $1$, \ldots, $6$ \\
    $\tan\beta$ & \lstinline|MINPAR[3]| & \lstinline|TB| & $\mathbb{R}_{>0}$ \\
    $m_{12}^2$ & \lstinline|MINPAR[18]| & \lstinline|m122| & $\mathbb{R}$ \\
    $\zeta_u$ & \lstinline|MINPAR[21]| & \lstinline|zetau| & $\mathbb{R}$ \\
    $\zeta_d$ & \lstinline|MINPAR[22]| & \lstinline|zetad| & $\mathbb{R}$ \\
    $\zeta_l$ & \lstinline|MINPAR[23]| & \lstinline|zetal| & $\mathbb{R}$ \\
    $\Delta_u$ & \lstinline|GM2CalcTHDMDeltauInput| & \lstinline|Deltau| & $\mathbb{R}^{3\times 3}$ \\
    $\Delta_d$ & \lstinline|GM2CalcTHDMDeltadInput| & \lstinline|Deltad| & $\mathbb{R}^{3\times 3}$ \\
    $\Delta_l$ & \lstinline|GM2CalcTHDMDeltalInput| & \lstinline|Deltal| & $\mathbb{R}^{3\times 3}$ \\
    $\Pi_u$ & \lstinline|GM2CalcTHDMPiuInput| & \lstinline|Piu| & $\mathbb{R}^{3\times 3}$ \\
    $\Pi_d$ & \lstinline|GM2CalcTHDMPidInput| & \lstinline|Pid| & $\mathbb{R}^{3\times 3}$ \\
    $\Pi_l$ & \lstinline|GM2CalcTHDMPilInput| & \lstinline|Pil| & $\mathbb{R}^{3\times 3}$ \\
    \bottomrule
  \end{tabular}
  \label{tab:basis2}
\end{table}
The ``mass basis'' input parameters are read from the
\lstinline|MINPAR| and \lstinline|MASS| blocks for compatibility with
\THDMC\ \cite{Eriksson:2009ws}.  The following shows an example input
in the ``mass basis'' for the type II \THDM:
\begin{lstlisting}
Block MINPAR                  # model parameters
    3        3                # tan(beta)
   16        0                # lambda_6
   17        0                # lambda_7
   18        40000            # m_{12}^2
   20        0.999            # sin(beta - alpha)
   21        0                # zeta_u (only used if Yukawa type = 5)
   22        0                # zeta_d (only used if Yukawa type = 5)
   23        0                # zeta_l (only used if Yukawa type = 5)
   24        2                # Yukawa type (1, 2, 3, 4, 5 = aligned, 6 = general)
Block MASS                    # Higgs masses
   25        125              # mh, lightest CP-even Higgs
   35        400              # mH, heaviest CP-even Higgs
   36        420              # mA, CP-odd Higgs
   37        440              # mH+, charged Higgs
\end{lstlisting}
Unset parameters in the \lstinline|MINPAR| and \lstinline|MASS| blocks
are assumed to be zero, except for $\tan\beta$ which will raise an
error. Entry 24 of the \lstinline|MINPAR| is used to select the type
of \THDM. Specifically the integer values $1$, \ldots, $6$ for the
Yukawa type correspond to: $1$ = type I, $2$ = type II, $3$ = type X,
$4$ = type Y, $5$ = Flavour-Aligned, $6$ = general \THDM.  The input
entries 21, 22, and 23 in the \lstinline|MINPAR| block can be used to
set the values of $\zeta_u$, $\zeta_d$ and $\zeta_l$ in the
Flavour-Aligned \THDM, and are ignored for all other types.
Additional blocks for the general case are described below.

\paragraph{Gauge basis input parameters}

Alternatively, the input can be given in the ``gauge basis'' in a
format compatible with \THDMC\ \cite{Eriksson:2009ws}.  In the ``gauge
basis'' the \THDM\ parameters must be given in the \lstinline|MINPAR|
block.  The available ``gauge basis'' input parameters are listed in
\tabref{tab:basis2}.  The following shows an example input in the
``gauge basis'':
\begin{lstlisting}
Block MINPAR                  # model parameters in gauge basis
    3        3                # tan(beta)
   11        0.7              # lambda_1
   12        0.6              # lambda_2
   13        0.5              # lambda_3
   14        0.4              # lambda_4
   15        0.3              # lambda_5
   16        0.2              # lambda_6
   17        0.1              # lambda_7
   18        40000            # m_{12}^2
   21        0                # zeta_u (only used if Yukawa type = 5)
   22        0                # zeta_d (only used if Yukawa type = 5)
   23        0                # zeta_l (only used if Yukawa type = 5)
   24        2                # Yukawa type (1, 2, 3, 4, 5 = aligned, 6 = general)
\end{lstlisting}
Two parametrizations for the general \THDM\ are implemented, see
Eq.~\eqref{eq:rho}. The first option is a deviation from the \FATHDM\
(or types I, II, X, Y), parametrized by the additional matrices
$\Delta_f$. Thus, after choosing Yukawa type = 1, \ldots, 5, i.e.\
type I, II, X, Y and \FATHDM, the matrices $\Delta_u$, $\Delta_d$ and
$\Delta_l$ can be optionally given in the following dedicated blocks:
\begin{lstlisting}
Block GM2CalcTHDMDeltauInput
    1 1   0                    # Re(Delta_u(1,1))
    1 2   0                    # Re(Delta_u(1,2))
    1 3   0                    # Re(Delta_u(1,3))
    2 1   0                    # Re(Delta_u(2,1))
    2 2   0                    # Re(Delta_u(2,2))
    2 3   0                    # Re(Delta_u(2,3))
    3 1   0                    # Re(Delta_u(3,1))
    3 2   0                    # Re(Delta_u(3,2))
    3 3   0                    # Re(Delta_u(3,3))
Block GM2CalcTHDMDeltadInput
    1 1   0                    # Re(Delta_d(1,1))
    1 2   0                    # Re(Delta_d(1,2))
    1 3   0                    # Re(Delta_d(1,3))
    2 1   0                    # Re(Delta_d(2,1))
    2 2   0                    # Re(Delta_d(2,2))
    2 3   0                    # Re(Delta_d(2,3))
    3 1   0                    # Re(Delta_d(3,1))
    3 2   0                    # Re(Delta_d(3,2))
    3 3   0                    # Re(Delta_d(3,3))
Block GM2CalcTHDMDeltalInput
    1 1   0                    # Re(Delta_l(1,1))
    1 2   0                    # Re(Delta_l(1,2))
    1 3   0                    # Re(Delta_l(1,3))
    2 1   0                    # Re(Delta_l(2,1))
    2 2   0.1                  # Re(Delta_l(2,2))
    2 3   0                    # Re(Delta_l(2,3))
    3 1   0                    # Re(Delta_l(3,1))
    3 2   0                    # Re(Delta_l(3,2))
    3 3   0                    # Re(Delta_l(3,3))
\end{lstlisting}
The other option for the general \THDM\ corresponds to the $\Pi$ parametrization, implemented when setting Yukawa type = 6. In this case, the input parameters $\Delta_u$, $\Delta_d$ and $\Delta_l$ are
ignored and the real parts of the matrices
$\Pi_u$, $\Pi_d$ and $\Pi_l$ can be given in the following dedicated
blocks:
\begin{lstlisting}
Block GM2CalcTHDMPiuInput
    1 1   0                    # Re(Pi_u(1,1))
    1 2   0                    # Re(Pi_u(1,2))
    1 3   0                    # Re(Pi_u(1,3))
    2 1   0                    # Re(Pi_u(2,1))
    2 2   0                    # Re(Pi_u(2,2))
    2 3   0                    # Re(Pi_u(2,3))
    3 1   0                    # Re(Pi_u(3,1))
    3 2   0                    # Re(Pi_u(3,2))
    3 3   0                    # Re(Pi_u(3,3))
Block GM2CalcTHDMPidInput
    1 1   0                    # Re(Pi_d(1,1))
    1 2   0                    # Re(Pi_d(1,2))
    1 3   0                    # Re(Pi_d(1,3))
    2 1   0                    # Re(Pi_d(2,1))
    2 2   0                    # Re(Pi_d(2,2))
    2 3   0                    # Re(Pi_d(2,3))
    3 1   0                    # Re(Pi_d(3,1))
    3 2   0                    # Re(Pi_d(3,2))
    3 3   0                    # Re(Pi_d(3,3))
Block GM2CalcTHDMPilInput
    1 1   0                    # Re(Pi_l(1,1))
    1 2   0                    # Re(Pi_l(1,2))
    1 3   0                    # Re(Pi_l(1,3))
    2 1   0                    # Re(Pi_l(2,1))
    2 2   0.1                  # Re(Pi_l(2,2))
    2 3   0                    # Re(Pi_l(2,3))
    3 1   0                    # Re(Pi_l(3,1))
    3 2   0                    # Re(Pi_l(3,2))
    3 3   0                    # Re(Pi_l(3,3))
\end{lstlisting}
The input parameters $\Pi_u$, $\Pi_d$ and $\Pi_l$ are ignored when choosing Yukawa type
= 1, \ldots, 5, i.e.\ type I, II, X, Y and \FATHDM.

The SLHA-like interface described so far is very convenient and
intuitive to use and it does not require any prior knowledge of
programming languages to run \GMTCalc\ this way. The execution time
for this is also very short, around $5\unit{ms}$ per point on current
laptops.\footnote{Based a machine with an Intel(R) Core(TM) i7-5600U
  CPU @ 2.60GHz processor.}  Nevertheless even faster execution times
and easy interfacing with other existing calculators and sampling
algorithms are enabled through our C++ ($0.05\unit{ms}$), C
($0.05\unit{ms}$), \mathematica\ ($0.5\unit{ms}$) and \python\
($0.08\unit{ms}$) interfaces.  In the following subsections we describe
how to use each of these interfaces. 

\subsection{Running \GMTCalc\ from within C++}
\label{sec:c++_interface}

\GMTCalc\ provides a C++ programming interface, which allows for
calculating of $\amuBSM$ in the \THDM\ up to the two-loop level.  The
following C++ source code snippet shows a two-loop example calculation
in the \THDM\ of type II with input parameters defined in the ``mass
basis'', see \tabref{tab:basis}.
\begin{lstlisting}[language=c++]
#include "gm2calc/gm2_1loop.hpp"
#include "gm2calc/gm2_2loop.hpp"
#include "gm2calc/gm2_uncertainty.hpp"
#include "gm2calc/gm2_error.hpp"
#include "gm2calc/THDM.hpp"

#include <cstdio>

int main()
{
   // define THDM parameters in the mass basis
   gm2calc::thdm::Mass_basis basis;
   basis.yukawa_type = gm2calc::thdm::Yukawa_type::type_2;
   basis.mh = 125;                             // light CP-even Higgs mass
   basis.mH = 400;                             // heavy CP-even Higgs mass
   basis.mA = 420;                             // CP-odd Higgs mass
   basis.mHp = 440;                            // charged Higgs mass
   basis.sin_beta_minus_alpha = 0.999;         // sin(beta - alpha)
   basis.lambda_6 = 0;                         // lambda_6
   basis.lambda_7 = 0;                         // lambda_7
   basis.tan_beta = 3;                         // tan(beta)
   basis.m122 = 40000;                         // m_{12}^2 in GeV^2
   basis.zeta_u = 0;                           // zeta_u
   basis.zeta_d = 0;                           // zeta_d
   basis.zeta_l = 0;                           // zeta_l
   basis.Delta_u << 0, 0, 0, 0, 0, 0, 0, 0, 0; // Delta_u
   basis.Delta_d << 0, 0, 0, 0, 0, 0, 0, 0, 0; // Delta_d
   basis.Delta_l << 0, 0, 0, 0, 0, 0, 0, 0, 0; // Delta_l
   basis.Pi_u << 0, 0, 0, 0, 0, 0, 0, 0, 0;    // Pi_u
   basis.Pi_d << 0, 0, 0, 0, 0, 0, 0, 0, 0;    // Pi_d
   basis.Pi_l << 0, 0, 0, 0, 0, 0, 0, 0, 0;    // Pi_l

   // define SM parameters
   gm2calc::SM sm;
   sm.set_alpha_em_mz(1.0/128.94579);  // electromagnetic coupling
   sm.set_mu(2, 173.34);               // top quark mass
   sm.set_mu(1, 1.28);                 // charm quark mass
   sm.set_md(2, 4.18);                 // bottom quark mass
   sm.set_ml(2, 1.77684);              // tau lepton mass

   // define options to customize the calculation
   gm2calc::thdm::Config config;
   config.running_couplings = true;    // use running couplings

   try {
      // setup the THDM
      gm2calc::THDM model(basis, sm, config);

      // calculate a_mu up to (including) the 2-loop level
      const double amu = gm2calc::calculate_amu_1loop(model)
                       + gm2calc::calculate_amu_2loop(model);

      // calculate the uncertainty of the 2-loop a_mu
      const double delta_amu =
         gm2calc::calculate_uncertainty_amu_2loop(model);

      std::printf("amu = %g +- %g\n", amu, delta_amu);
   } catch (const gm2calc::Error& e) {
      std::printf("%s\n", e.what());
   }

   return 0;
}
\end{lstlisting}
This example source code can be compiled as follows (assuming
\GMTCalc\ has been compiled to a shared library
\lstinline|libgm2calc.so| on a UNIX-like operating system with
\lstinline|g++| installed):
\begin{lstlisting}[language=bash]
g++ -I${GM2CALC_DIR}/include/ -I${EIGEN_DIR} example.cpp ${GM2CALC_DIR}/build/lib/libgm2calc.so
\end{lstlisting}
Here \lstinline|example.cpp| is the file that contains the above
listed source code.  The variable \lstinline|GM2CALC_DIR| contains the
path to the \GMTCalc\ root directory and \lstinline|EIGEN_DIR|
contains the path to the Eigen library header files.  Running the
created executable \lstinline|a.out| yields
\begin{lstlisting}[language=bash]
$ ./a.out
amu = 1.67323e-11 +- 3.3616e-12
\end{lstlisting}
In line~12 of the example source code an object of type
\lstinline|Mass_basis| is created, which contains the input parameters
in the mass basis, see \tabref{tab:basis}.  The mass basis input
parameters are set in lines 13--31.  Note that the Yukawa type is set
to type II in line~13, which implies that given values of $\zeta_f$
are ignored and internally fixed to the values given in
\tabref{tab:zeta}.  
Additionally the inputs $\Pi_f$ and $\Delta_f$ are unused for this type, and ignored.  
In line~34 an object of type \lstinline|SM| is created that contains
all \SM\ input parameters.  The \SM\ input parameters are set to
reasonable default values from the PDG \cite{Zyla:2020zbs}.  In
lines~35--39 the values for $\aem(m_Z)$, $m_t$, $m_c^\MS(2\GeV)$,
$m_b^\MS(m_b^\MS)$ and $m_\tau$ are set to specific values.
In line~42 an object of type \lstinline|Config| is created, which
contains the options to customize the calculation of $\amuBSM$ and
$\Damu$.  In line~43 the ``running masses'' scheme is chosen, see
\secref{sec:running_couplings}.
In line~47 the \THDM\ model is created, given the \THDM\ and \SM\ input
parameters and the configuration options defined above.  The value of
$\amuBSM = \amu^\ol + \amu^\tl$ is calculated in lines~50--51.  The
corresponding uncertainty $\Damu$ is calculated in line~54--55.  The
values $\amuBSM$ and $\Damu$ are printed in line~57.
Note that the \THDM\ model should be created within a \lstinline|try|
block, because the constructor of the \THDM\ class throws an exception
if a physical problem occurs (e.g.\ a tachyon) or an input parameter
has been set to an invalid value, see \tabref{tab:basis2}.

Alternatively, the \THDM\ input parameters can be given in the ``gauge
basis'', i.e.\ in terms of the Lagrangian parameters, see
\tabref{tab:basis}.  The following C++ source code snippet shows a
corresponding example two-loop calculation in the \THDM\ of type II,
where the input parameters are defined in the ``gauge basis''.
\begin{lstlisting}[language=c++]
#include "gm2calc/gm2_1loop.hpp"
#include "gm2calc/gm2_2loop.hpp"
#include "gm2calc/gm2_uncertainty.hpp"
#include "gm2calc/gm2_error.hpp"
#include "gm2calc/THDM.hpp"

#include <cstdio>

int main()
{
   // define THDM parameters in the gauge basis
   gm2calc::thdm::Gauge_basis basis;
   basis.yukawa_type = gm2calc::thdm::Yukawa_type::type_2;
   basis.lambda << 0.7, 0.6, 0.5, 0.4, 0.3, 0.2, 0.1;   // lambda_{1,...,7}
   basis.tan_beta = 3;                                  // tan(beta)
   basis.m122 = 40000;                                  // m_{12}^2 in GeV^2
   basis.zeta_u = 0;                                    // zeta_u
   basis.zeta_d = 0;                                    // zeta_d
   basis.zeta_l = 0;                                    // zeta_l
   basis.Delta_u << 0, 0, 0, 0, 0, 0, 0, 0, 0;          // Delta_u
   basis.Delta_d << 0, 0, 0, 0, 0, 0, 0, 0, 0;          // Delta_d
   basis.Delta_l << 0, 0, 0, 0, 0, 0, 0, 0, 0;          // Delta_l
   basis.Pi_u << 0, 0, 0, 0, 0, 0, 0, 0, 0;             // Pi_u
   basis.Pi_d << 0, 0, 0, 0, 0, 0, 0, 0, 0;             // Pi_d
   basis.Pi_l << 0, 0, 0, 0, 0, 0, 0, 0, 0;             // Pi_l

   // define SM parameters
   gm2calc::SM sm;
   sm.set_alpha_em_mz(1.0/128.94579);  // electromagnetic coupling
   sm.set_mu(2, 173.34);               // top quark mass
   sm.set_mu(1, 1.28);                 // charm quark mass
   sm.set_md(2, 4.18);                 // bottom quark mass
   sm.set_ml(2, 1.77684);              // tau lepton mass

   // define options to customize the calculation
   gm2calc::thdm::Config config;
   config.running_couplings = true;    // use running couplings

   try {
      // setup the THDM
      gm2calc::THDM model(basis, sm, config);

      // calculate a_mu up to (including) the 2-loop level
      const double amu = gm2calc::calculate_amu_1loop(model)
                       + gm2calc::calculate_amu_2loop(model);

      // calculate the uncertainty of the 2-loop a_mu
      const double delta_amu =
         gm2calc::calculate_uncertainty_amu_2loop(model);

      std::printf("amu = %g +- %g\n", amu, delta_amu);
   } catch (const gm2calc::Error& e) {
      std::printf("%s\n", e.what());
   }

   return 0;
}
\end{lstlisting}
In line~12 an object of type \lstinline|Gauge_basis| is created, which
is filled with the gauge basis input parameters in lines~13--25.  With
the defined gauge basis input parameters the calculation of $\amuBSM$
and $\Damu$ continues as in the mass basis example above.

\subsection{Running \GMTCalc\ from within C}

Alternatively to the C++ programming interface detailed in
\secref{sec:c++_interface}, \GMTCalc\ also provides a C programming
interface.  The following C source code snippet shows a two-loop
example calculation in the \THDM\ of type II with input parameters
defined in the ``mass basis'', see \tabref{tab:basis}.
\begin{lstlisting}[language=c]
#include "gm2calc/gm2_1loop.h"
#include "gm2calc/gm2_2loop.h"
#include "gm2calc/gm2_uncertainty.h"
#include "gm2calc/THDM.h"
#include "gm2calc/SM.h"

#include <stdio.h>

int main()
{
   /* define THDM parameters in the mass basis */
   const gm2calc_THDM_mass_basis basis = {
      .yukawa_type = gm2calc_THDM_type_2,       /* Yukawa type */
      .mh = 125,                                /* light CP-even Higgs mass */
      .mH = 400,                                /* heavy CP-even Higgs mass */
      .mA = 420,                                /* CP-odd Higgs mass */
      .mHp = 440,                               /* charged Higgs mass */
      .sin_beta_minus_alpha = 0.999,            /* sin(beta - alpha) */
      .lambda_6 = 0,                            /* lambda_6 */
      .lambda_7 = 0,                            /* lambda_7 */
      .tan_beta = 3,                            /* tan(beta) */
      .m122 = 40000,                            /* m_{12}^2 in GeV^2 */
      .zeta_u = 0,                              /* zeta_u */
      .zeta_d = 0,                              /* zeta_d */
      .zeta_l = 0,                              /* zeta_l */
      .Delta_u = { {0,0,0}, {0,0,0}, {0,0,0} }, /* Re(Delta_u(i,k)) */
      .Delta_d = { {0,0,0}, {0,0,0}, {0,0,0} }, /* Re(Delta_d(i,k)) */
      .Delta_l = { {0,0,0}, {0,0,0}, {0,0,0} }, /* Re(Delta_l(i,k)) */
      .Pi_u = { {0,0,0}, {0,0,0}, {0,0,0} },    /* Re(Pi_u(i,k)) */
      .Pi_d = { {0,0,0}, {0,0,0}, {0,0,0} },    /* Re(Pi_d(i,k)) */
      .Pi_l = { {0,0,0}, {0,0,0}, {0,0,0} }     /* Re(Pi_l(i,k)) */
   };

   /* define SM parameters */
   gm2calc_SM sm;
   gm2calc_sm_set_to_default(&sm);
   sm.alpha_em_mz = 1.0/128.94579;  /* electromagnetic coupling */
   sm.mu[2] = 173.34;               /* top quark mass */
   sm.mu[1] = 1.28;                 /* charm quark mass */
   sm.md[2] = 4.18;                 /* bottom quark mass */
   sm.ml[2] = 1.77684;              /* tau lepton mass */

   /* calculation settings */
   gm2calc_THDM_config config;
   gm2calc_thdm_config_set_to_default(&config);

   /* setup the THDM */
   gm2calc_THDM* model = 0;
   gm2calc_error error = gm2calc_thdm_new_with_mass_basis(&model, &basis, &sm, &config);

   if (error == gm2calc_NoError) {
      /* calculate a_mu up to (including) the 2-loop level */
      const double amu = gm2calc_thdm_calculate_amu_1loop(model)
                       + gm2calc_thdm_calculate_amu_2loop(model);

      /* calculate the uncertainty of the 2-loop a_mu */
      const double delta_amu =
         gm2calc_thdm_calculate_uncertainty_amu_2loop(model);

      printf("amu = %g +- %g\n", amu, delta_amu);
   } else {
      printf("Error: %s\n", gm2calc_error_str(error));
   }

   gm2calc_thdm_free(model);

   return 0;
}
\end{lstlisting}
This example source code can be compiled as follows (assuming
\GMTCalc\ has been compiled to a shared library
\lstinline|libgm2calc.so| on a UNIX-like operating system with
\lstinline|gcc| installed):
\begin{lstlisting}[language=bash]
gcc -I${GM2CALC_DIR}/include/ example.c ${GM2CALC_DIR}/build/lib/libgm2calc.so
\end{lstlisting}
Here \lstinline|example.c| is the file that contains the above
listed source code.

The C example source code is very similar to the mass basis C++
example.  In line~12 an object of type
\lstinline|gm2calc_THDM_mass_basis| is created and filled with the
mass basis input parameters in lines~13--32.  The Yukawa type of the
model is defined in line~13 to be type II.
In line~35 an object of type \lstinline|gm2calc_SM| is created, which
contains the \SM\ input parameters.  The \SM\ input parameters are set to
their default values in line~36.  In lines~37--41 the values of
$\aem(m_Z)$, $m_t$, $m_c^\MS(2\GeV)$, $m_b^\MS(m_b^\MS)$ and
$m_\tau$ are set to specific values.
In line~44 a config object of type \lstinline|gm2calc_THDM_config| is
created which contains options to customize the calculation.  These
options are set to default values in line~45.
In line~48 a null-pointer to a \THDM\ model of type
\lstinline|gm2calc_THDM| is created.  In line~49 the \THDM\ model is
created and the pointer is set to point to the model.  If an error
occurrs, the pointer is set to 0 and the returned \lstinline|error|
variable is set to a value that is not \lstinline|gm2calc_NoError|.
If no error has occurred, the example continues to calculate $\amuBSM$
and $\Damu$ in lines~53--58, respectively.  The result is printed in
line~60.  In line~65 the memory reserved for the \THDM\ model is freed.

Alternatively, the \THDM\ input parameters can be given in the ``gauge
basis'', similarly to the gauge basis C++ example.  The following C
source code snippet shows a corresponding example two-loop calculation
in the \THDM\ of type II, where the input parameters are defined in the
``gauge basis''.
\begin{lstlisting}[language=c]
#include "gm2calc/gm2_1loop.h"
#include "gm2calc/gm2_2loop.h"
#include "gm2calc/gm2_uncertainty.h"
#include "gm2calc/THDM.h"
#include "gm2calc/SM.h"

#include <stdio.h>

int main()
{
   /* define THDM parameters in the gauge basis */
   const gm2calc_THDM_gauge_basis basis = {
      .yukawa_type = gm2calc_THDM_type_2,              /* Yukawa type */
      .lambda = { 0.7, 0.6, 0.5, 0.4, 0.3, 0.2, 0.1 }, /* lambda_i */
      .tan_beta = 3,                                   /* tan(beta) */
      .m122 = 40000,                                   /* m_{12}^2 in GeV^2 */
      .zeta_u = 0,                                     /* zeta_u */
      .zeta_d = 0,                                     /* zeta_d */
      .zeta_l = 0,                                     /* zeta_l */
      .Delta_u = { {0,0,0}, {0,0,0}, {0,0,0} },        /* Re(Delta_u(i,k)) */
      .Delta_d = { {0,0,0}, {0,0,0}, {0,0,0} },        /* Re(Delta_u(i,k)) */
      .Delta_l = { {0,0,0}, {0,0,0}, {0,0,0} },        /* Re(Delta_u(i,k)) */
      .Pi_u = { {0,0,0}, {0,0,0}, {0,0,0} },           /* Re(Pi_u(i,k)) */
      .Pi_d = { {0,0,0}, {0,0,0}, {0,0,0} },           /* Re(Pi_d(i,k)) */
      .Pi_l = { {0,0,0}, {0,0,0}, {0,0,0} }            /* Re(Pi_l(i,k)) */
   };

   /* define SM parameters */
   gm2calc_SM sm;
   gm2calc_sm_set_to_default(&sm);
   sm.alpha_em_mz = 1.0/128.94579;  /* electromagnetic coupling */
   sm.mu[2] = 173.34;               /* top quark mass */
   sm.mu[1] = 1.28;                 /* charm quark mass */
   sm.md[2] = 4.18;                 /* bottom quark mass */
   sm.ml[2] = 1.77684;              /* tau lepton mass */

   /* calculation settings */
   gm2calc_THDM_config config;
   gm2calc_thdm_config_set_to_default(&config);

   /* setup the THDM */
   gm2calc_THDM* model = 0;
   gm2calc_error error = gm2calc_thdm_new_with_gauge_basis(&model, &basis, &sm, &config);

   if (error == gm2calc_NoError) {
      /* calculate a_mu up to (including) the 2-loop level */
      const double amu = gm2calc_thdm_calculate_amu_1loop(model)
                       + gm2calc_thdm_calculate_amu_2loop(model);

      /* calculate the uncertainty of the 2-loop a_mu */
      const double delta_amu =
         gm2calc_thdm_calculate_uncertainty_amu_2loop(model);

      printf("amu = %g +- %g\n", amu, delta_amu);
   } else {
      printf("Error: %s\n", gm2calc_error_str(error));
   }

   gm2calc_thdm_free(model);

   return 0;
}
\end{lstlisting}
In line~12 an object of type \lstinline|gm2calc_THDM_gauge_basis| is
created, which is filled with the gauge basis input parameters in
lines~13--26.  In line~43 the \THDM\ model is created, using the gauge
basis input parameters.  The calculation of $\amuBSM$ and $\Damu$ is
performed in lines~47--52, as in the ``mass basis'' example above.

\subsection{Running \GMTCalc\ from within \mathematica}

\GMTCalc\ can be run from within \mathematica\ using the
\lstinline|MathLink| interface.  The following source code snippet shows
an example calculation of $\amuBSM$ and its uncertainty at the two-loop
level using input parameters given in the ``gauge basis''.
\begin{lstlisting}[language=mathematica]
Install["bin/gm2calc.mx"];

(* calculation settings *)
GM2CalcSetFlags[
    loopOrder -> 2,
    runningCouplings -> True];

(* set SM parameters *)
GM2CalcSetSMParameters[
    alpha0 -> 0.00729735,      (* alpha_em in Thompson limit *)
    alphaMZ -> 0.0077552,      (* alpha_em(MZ) *)
    alphaS -> 0.1184,          (* alpha_s *)
    MhSM -> 125.09,            (* SM Higgs boson pole mass *)
    MW -> 80.385,              (* W boson pole mass *)
    MZ -> 91.1876,             (* Z boson pole mass *)
    MT -> 173.34,              (* top quark pole mass *)
    mcmc -> 1.28,              (* charm quark MS-bar mass mc at Q = mc *)
    mu2GeV -> 0.0022,          (* up quark MS-bar mass at Q = 2 GeV *)
    mbmb -> 4.18,              (* bottom quark MS-bar mass mb at Q = mb *)
    ms2GeV -> 0.096,           (* strange quark MS-bar mass at Q = 2 GeV *)
    md2GeV -> 0.0047,          (* down quark MS-bar mass at Q = 2 GeV *)
    ML -> 1.777,               (* tau lepton pole mass *)
    MM -> 0.1056583715,        (* muon pole mass *)
    ME -> 0.000510998928,      (* electron pole mass *)
    Mv1 -> 0,                  (* lightest neutrino mass *)
    Mv2 -> 0,                  (* 2nd lightest neutrino mass *)
    Mv3 -> 0,                  (* heaviest neutrino mass *)
    CKM -> IdentityMatrix[3]   (* CKM matrix *)
];

(* calculate amu using the gauge basis input parameters *)
result = {amu, Damu} /. GM2CalcAmuTHDMGaugeBasis[
    yukawaType        -> 2,                    (* Yukawa type (1,...,6) *)
    lambda            -> { 0.7, 0.6, 0.5, 0.4, 0.3, 0.2, 0.1 }, (* lambda_{1,...,7} *)
    TB                -> 3,                    (* tan(beta) = v2/v1 *)
    m122              -> 200^2,                (* m_{12}^2 *)
    zetau             -> 0,                    (* zeta_u *)
    zetad             -> 0,                    (* zeta_d *)
    zetal             -> 0,                    (* zeta_l *)
    Deltau            -> 0 IdentityMatrix[3],  (* Delta_u *)
    Deltad            -> 0 IdentityMatrix[3],  (* Delta_d *)
    Deltal            -> 0 IdentityMatrix[3],  (* Delta_l *)
    Piu               -> 0 IdentityMatrix[3],  (* Pi_u *)
    Pid               -> 0 IdentityMatrix[3],  (* Pi_d *)
    Pil               -> 0 IdentityMatrix[3]   (* Pi_l *)
  ];

Print[result];
\end{lstlisting}
In line~1 \GMTCalc's \lstinline|MathLink| executable
\lstinline|bin/gm2calc.mx|, which is created when building \GMTCalc,
is loaded into the \mathematica\ session.  In lines~4--6 two
configuration options to customize the calculation are set: The
calculation shall be performed at the two-loop level using the
``running masses'' scheme defined in \secref{sec:running_couplings}.
In lines~9--29 the \SM\ input parameters are defined.  Unset parameters
are set to reasonable default values, see
\lstinline|Options[GM2CalcSetSMParameters]|.  In lines~32--46 the
values of $\amuBSM$ and $\Damu$ are calculated using the function
\lstinline|GM2CalcAmuTHDMGaugeBasis|, which takes the gauge basis
input parameters as arguments, see \tabref{tab:basis2}.  The result is
printed in line~48.

Alternatively, the calculation can be performed using input parameters
given in the ``mass basis''.  The following source code snippet shows
an example calculation of $\amuBSM$ and its uncertainty at the two-loop
level using input parameters given in the ``mass basis''.
\begin{lstlisting}[language=mathematica]
Install["bin/gm2calc.mx"];

(* calculation settings *)
GM2CalcSetFlags[
    loopOrder -> 2,
    runningCouplings -> True];

(* set SM parameters *)
GM2CalcSetSMParameters[
    alpha0 -> 0.00729735,      (* alpha_em in Thompson limit *)
    alphaMZ -> 0.0077552,      (* alpha_em(MZ) *)
    alphaS -> 0.1184,          (* alpha_s *)
    MhSM -> 125.09,            (* SM Higgs boson pole mass *)
    MW -> 80.385,              (* W boson pole mass *)
    MZ -> 91.1876,             (* Z boson pole mass *)
    MT -> 173.34,              (* top quark pole mass *)
    mcmc -> 1.28,              (* charm quark MS-bar mass mc at Q = mc *)
    mu2GeV -> 0.0022,          (* up quark MS-bar mass at Q = 2 GeV *)
    mbmb -> 4.18,              (* bottom quark MS-bar mass mb at Q = mb *)
    ms2GeV -> 0.096,           (* strange quark MS-bar mass at Q = 2 GeV *)
    md2GeV -> 0.0047,          (* down quark MS-bar mass at Q = 2 GeV *)
    ML -> 1.777,               (* tau lepton pole mass *)
    MM -> 0.1056583715,        (* muon pole mass *)
    ME -> 0.000510998928,      (* electron pole mass *)
    Mv1 -> 0,                  (* lightest neutrino mass *)
    Mv2 -> 0,                  (* 2nd lightest neutrino mass *)
    Mv3 -> 0,                  (* heaviest neutrino mass *)
    CKM -> IdentityMatrix[3]   (* CKM matrix *)
];

(* calculate amu using the mass basis input parameters *)
result = {amu, Damu} /. GM2CalcAmuTHDMMassBasis[
    yukawaType        -> 2,                    (* Yukawa type (1,...,6) *)
    Mhh               -> { 125, 400 },         (* CP-even Higgs boson masses *)
    MAh               -> 420,                  (* CP-odd Higgs boson mass *)
    MHp               -> 440,                  (* charged Higgs boson mass *)
    sinBetaMinusAlpha -> 0.999,                (* sin(beta - alpha) *)
    lambda6           -> 0,                    (* lambda_6 *)
    lambda7           -> 0,                    (* lambda_7 *)
    TB                -> 3,                    (* tan(beta) = v2/v1 *)
    m122              -> 200^2,                (* m_{12}^2 *)
    zetau             -> 0,                    (* zeta_u *)
    zetad             -> 0,                    (* zeta_d *)
    zetal             -> 0,                    (* zeta_l *)
    Deltau            -> 0 IdentityMatrix[3],  (* Delta_u *)
    Deltad            -> 0 IdentityMatrix[3],  (* Delta_d *)
    Deltal            -> 0 IdentityMatrix[3],  (* Delta_l *)
    Piu               -> 0 IdentityMatrix[3],  (* Pi_u *)
    Pid               -> 0 IdentityMatrix[3],  (* Pi_d *)
    Pil               -> 0 IdentityMatrix[3]   (* Pi_l *)
  ];

Print[result];
\end{lstlisting}
The calculation of $\amuBSM$ and $\Damu$ is performed in lines~32--51
with the function \lstinline|GM2CalcAmuTHDMMassBasis|, which takes the
mass basis input parameters as arguments, see \tabref{tab:basis2}.
The result is printed in line~53.

\subsection{Running \GMTCalc\ from within \python}
\label{sec:python_interface}

Newly implemented in \GMTCalc\ 2.0.0 is the ability to interface with \python\ using the package \lstinline|cppyy|.  An example calculation using the interface is shown in the code snippet below, working in the ``mass basis''.
\begin{lstlisting}[language=python]
#!/usr/bin/env python

from __future__ import print_function
from gm2_python_interface import *

cppyy.include(os.path.join("gm2calc","gm2_1loop.hpp"))
cppyy.include(os.path.join("gm2calc","gm2_2loop.hpp"))
cppyy.include(os.path.join("gm2calc","gm2_uncertainty.hpp"))
cppyy.include(os.path.join("gm2calc","gm2_error.hpp"))
cppyy.include(os.path.join("gm2calc","SM.hpp"))
cppyy.include(os.path.join("gm2calc","THDM.hpp"))

cppyy.load_library("libgm2calc")

# Load data types
from cppyy.gbl import Eigen
from cppyy.gbl import gm2calc
from cppyy.gbl.gm2calc import SM
from cppyy.gbl.gm2calc import THDM
from cppyy.gbl.gm2calc import Error

# define THDM parameters in the mass basis
basis = gm2calc.thdm.Mass_basis()
basis.yukawa_type = gm2calc.thdm.Yukawa_type.type_2
basis.mh = 125.                            # light CP-even Higgs mass
basis.mH = 400.                            # heavy CP-even Higgs mass
basis.mA = 420.                            # CP-odd Higgs mass
basis.mHp = 440.                           # charged Higgs mass
basis.sin_beta_minus_alpha = 0.999         # sin(beta - alpha)
basis.lambda_6 = 0.                        # lambda_6
basis.lambda_7 = 0.                        # lambda_7
basis.tan_beta = 3.                        # tan(beta)
basis.m122 = 40000.                        # m_{12}^2 in GeV^2
basis.zeta_u = 0.                          # zeta_u
basis.zeta_d = 0.                          # zeta_d
basis.zeta_l = 0.                          # zeta_l
basis.Delta_u = Eigen.Matrix3d().setZero() # Delta_u
basis.Delta_d = Eigen.Matrix3d().setZero() # Delta_d
basis.Delta_l = Eigen.Matrix3d().setZero() # Delta_l
basis.Pi_u = Eigen.Matrix3d().setZero()    # Pi_u
basis.Pi_d = Eigen.Matrix3d().setZero()    # Pi_d
basis.Pi_l = Eigen.Matrix3d().setZero()    # Pi_l
# define SM parameters
sm = gm2calc.SM()
sm.set_alpha_em_mz(1.0/128.94579)  # electromagnetic coupling
sm.set_mu(2, 173.34)               # top quark mass
sm.set_mu(1, 1.28)                 # charm quark mass
sm.set_md(2, 4.18)                 # bottom quark mass
sm.set_ml(2, 1.77684)              # tau lepton mass
# define options to customize the calculation
config = gm2calc.thdm.Config()
config.running_couplings = True;   # use running couplings

try:
    # setup the THDM
    model = gm2calc.THDM(basis,sm,config)
    # calculate a_mu up to (including) the 2-loop level
    amu = gm2calc.calculate_amu_1loop(model) + gm2calc.calculate_amu_2loop(model)
    # calculate the uncertainty of the 2-loop a_mu
    delta_amu = gm2calc.calculate_uncertainty_amu_2loop(model)
    print("amu =",amu,"+-",delta_amu)
except gm2calc.Error as e:
    print(e.what())

\end{lstlisting}
Note similarity between the above code and the C++ and C interfaces.
Line~3 is to make sure that this example which is written using
\python\ 3-style \lstinline|print| functions can still work in
\python\ 2.  Line~4 imports the interface script
\lstinline|gm2_python_interface| which loads the \lstinline|cppyy| and
\lstinline|os| packages, as well as the header and library
locations.  This interface file is originally in the \lstinline|src| subdirectory.  
After performing
\begin{lstlisting}[language=bash]
cmake -DBUILD_SHARED_LIBS=ON .. 
\end{lstlisting}
the interface script will be copied into the subdirectory
\lstinline|bin|, and it will be filled with the path information 
for the \GMTCalc\ headers, library, and the \lstinline|Eigen3| 
path.  The interface can be imported from there by other \python\ scripts, 
or moved to an appropriate location where the user has their own \python\ scripts.  
Lines~6--11 load the relevant header
files, and line~13 loads the \GMTCalc\ shared library.  In
lines~16--20 the necessary namespaces from C++ are loaded into
\python.  In line~23 the \THDM\ \lstinline|Mass_basis|
object is initialized, while on line~24 the \THDM\ is specified to be
type II.  Lines~25--36 involve setting the values for simple
attributes in the basis.  Lines~36--42 assign values to the
\lstinline|Eigen::Matrix| attributes, however since these are meant to
be ignored, they are just set to $0$.  Lines~44--49 initialize an
\lstinline|SM| object and ensures it has the appropriate parameters.
Line 51 initializes the \lstinline|config| object, which is used to
flag the use of running coupling in the next line.  Line~56
initializes a \lstinline|THDM| object using the
\lstinline|Mass_basis|, \lstinline|SM|, and \lstinline|Config|
information.  Lines~58--63 prints out the values of $\amuBSM$ and
$\Damu$ which are calculated using the interface functions
\lstinline|calculate_amu_1loop|, \lstinline|calculate_amu_2loop|, and
\lstinline|calculate_uncertainty_amu_2loop|.  Alternatively an error
message will be printed out on line~63 should a problem arise.

Another example of the \python\ interface is shown below, this time using the ``gauge basis'':
\begin{lstlisting}[language=python]
#!/usr/bin/env python

from __future__ import print_function
from gm2_python_interface import *

cppyy.include(os.path.join("gm2calc","gm2_1loop.hpp"))
cppyy.include(os.path.join("gm2calc","gm2_2loop.hpp"))
cppyy.include(os.path.join("gm2calc","gm2_uncertainty.hpp"))
cppyy.include(os.path.join("gm2calc","gm2_error.hpp"))
cppyy.include(os.path.join("gm2calc","SM.hpp"))
cppyy.include(os.path.join("gm2calc","THDM.hpp"))

cppyy.load_library("libgm2calc")

# Load data types
from cppyy.gbl import Eigen
from cppyy.gbl import gm2calc
from cppyy.gbl.gm2calc import SM
from cppyy.gbl.gm2calc import THDM
from cppyy.gbl.gm2calc import Error

# define THDM parameters in the mass basis
basis = gm2calc.thdm.Gauge_basis()
basis.yukawa_type = gm2calc.thdm.Yukawa_type.type_2
# lambda is a reserved keyword, so we have to get creative
Matrix7d = Eigen.Matrix('double',7,1)
lam = Matrix7d().setZero()
lam[0] = 0.7
lam[1] = 0.6
lam[2] = 0.5
lam[3] = 0.4
lam[4] = 0.3
lam[5] = 0.2
lam[6] = 0.1
basis.__setattr__('lambda',lam)            # lambda_{1,...,7}
basis.tan_beta = 3.                        # tan(beta)
basis.m122 = 40000.                        # m_{12}^2 in GeV^2
basis.zeta_u = 0.                          # zeta_u
basis.zeta_d = 0.                          # zeta_d
basis.zeta_l = 0.                          # zeta_l
basis.Delta_u = Eigen.Matrix3d().setZero() # Delta_u
basis.Delta_d = Eigen.Matrix3d().setZero() # Delta_d
basis.Delta_l = Eigen.Matrix3d().setZero() # Delta_l
basis.Pi_u = Eigen.Matrix3d().setZero()    # Pi_u
basis.Pi_d = Eigen.Matrix3d().setZero()    # Pi_d
basis.Pi_l = Eigen.Matrix3d().setZero()    # Pi_l
# define SM parameters
sm = gm2calc.SM()
sm.set_alpha_em_mz(1.0/128.94579)  # electromagnetic coupling
sm.set_mu(2, 173.34)               # top quark mass
sm.set_mu(1, 1.28)                 # charm quark mass
sm.set_md(2, 4.18)                 # bottom quark mass
sm.set_ml(2, 1.77684)              # tau lepton mass
# define options to customize the calculation
config = gm2calc.thdm.Config()
config.running_couplings = True;   # use running couplings

try:
    # setup the THDM
    model = gm2calc.THDM(basis,sm,config)
    # calculate a_mu up to (including) the 2-loop level
    amu = gm2calc.calculate_amu_1loop(model) + gm2calc.calculate_amu_2loop(model)
    # calculate the uncertainty of the 2-loop a_mu
    delta_amu = gm2calc.calculate_uncertainty_amu_2loop(model)
    print("amu =",amu,"+-",delta_amu)
except gm2calc.Error as e:
    print(e.what())

\end{lstlisting}
In line~23 we instead initialize a \lstinline|Gauge_basis| object.  To define the attribute \lstinline|lambda|, we need to circumvent \python's reserved keywords.  This is done by defining a $7\times1$ \lstinline|Eigen::Matrix| in line~26.  This \lstinline|Matrix| is initialized to $0$ before assigned the appropriate entires elementwise on lines~28--34.  Then the method \lstinline|__setattr__| can be used to interface the values to the C++ code.  Then the other values can be defined on lines~36--53, and finally the result for $\amuBSM$ is printed on line~65.

\section{Applications}
\label{sec:applications}

\subsection{Parameter scan in the type II and X models}

\begin{figure}[tb]
  \centering
  \includegraphics[width=0.49\textwidth]{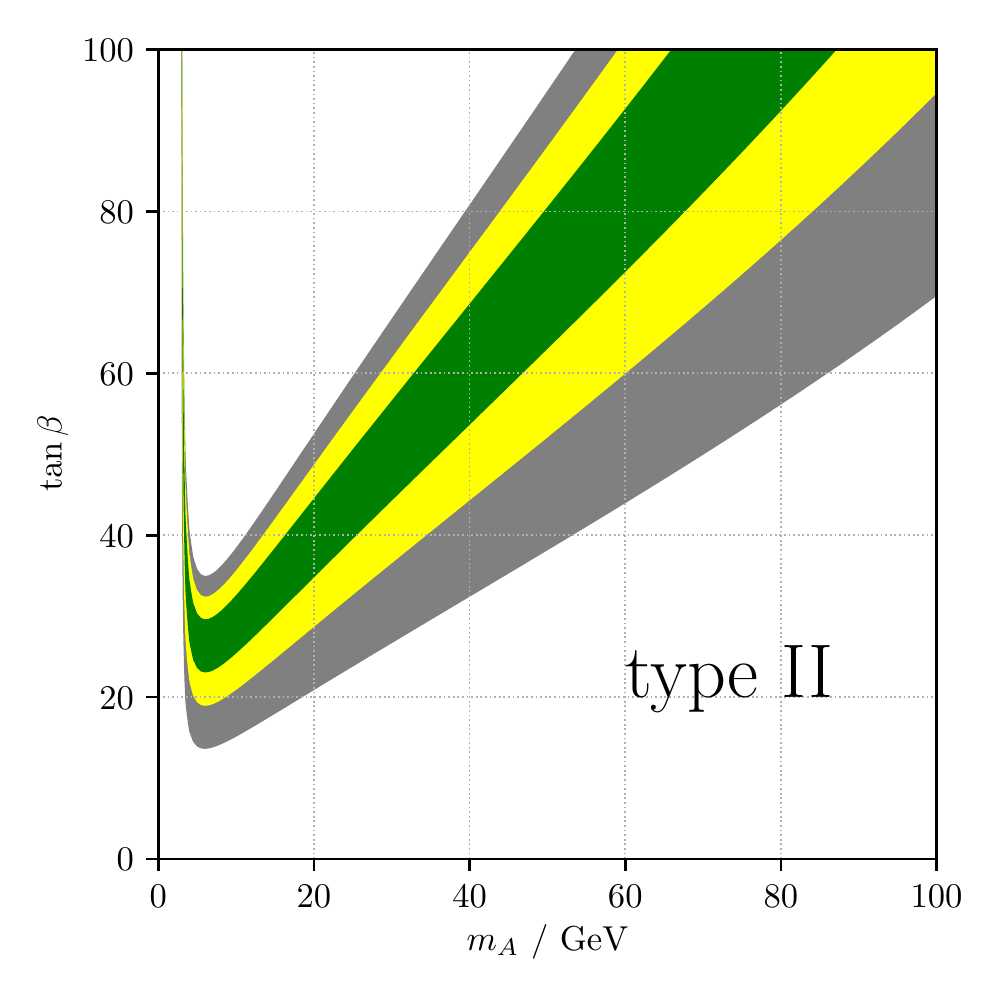}\hfill
  \includegraphics[width=0.49\textwidth]{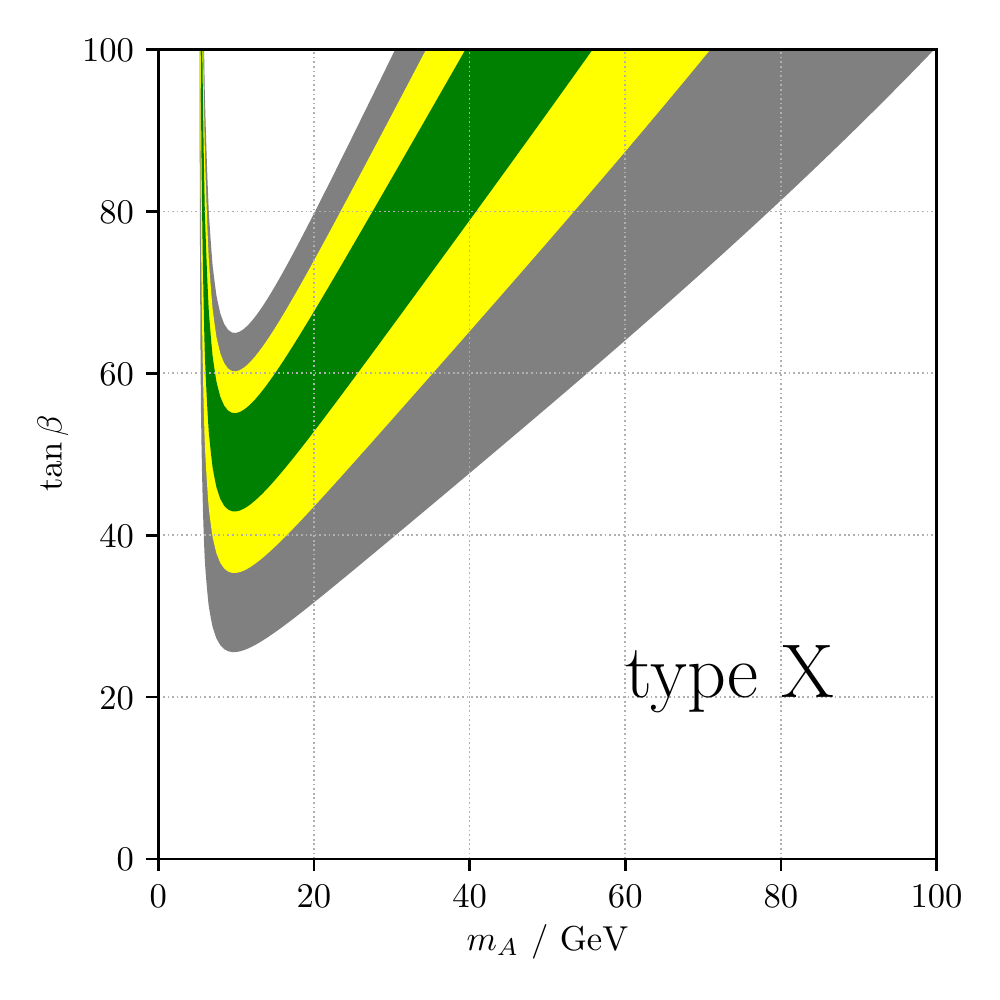}
  \caption{Two-loop prediction of $\amuBSM$ in the \THDM\ of type
    II (left) and type X (right) as a function of $\tan\beta$ and
    $m_A$ with $m_h=126\GeV$, $m_H=m_{H^\pm}=200\GeV$,
    $\sin(\beta-\alpha)=1$, $\lambda_6=\lambda_7=0$,
    $m_{12}^2=m_H^2/\tan\beta + (m_h^2 - \lambda_1 v^2)/\tan^3\beta$
    and $\lambda_1=\sqrt{4\pi}$.  In the green, yellow and gray
    regions the \THDM\ predicts the correct value of
    $\amuBSM = (25.1 \pm 5.9)\times 10^{-10}$ from Eq.\
    \eqref{eq:amuBSM} within one, two and three standard deviations,
    respectively.}
  \label{fig:mA-tb}
\end{figure}

As an application we perform a 2-dimensional parameter scan over $m_A$
and $\tan\beta$ for the type II and type X \THDM\ models,
similarly to Ref.\ \cite{Broggio:2014mna}.  However, in contrast to
Ref.\ \cite{Broggio:2014mna} we include the two-loop bosonic
contributions and use the updated value of
$\amuBSM = (25.1 \pm 5.9)\times 10^{-10}$ from Eq.\
\eqref{eq:amuBSM}.  The following C++ source code shows the program to
perform the scan.
\begin{lstlisting}[language=c++]
#include "gm2calc/gm2_1loop.hpp"
#include "gm2calc/gm2_2loop.hpp"
#include "gm2calc/gm2_error.hpp"
#include "gm2calc/THDM.hpp"

#include <cstdio>
#include <limits>

// Calculates amu in the mA-tan(beta) plane as in Fig.3 from
// arxiv:1409.3199.
double calc_amu(double mA, double tb, gm2calc::thdm::Yukawa_type yukawa_type)
{
   gm2calc::SM sm;
   const double v = sm.get_v();
   const double mh = 126, mH = 200;
   const double lambda_max = 3.5449077018110321; // Sqrt[4 Pi]
   const double lambda1 = lambda_max;

   gm2calc::thdm::Mass_basis basis;
   basis.yukawa_type = yukawa_type;
   basis.mh = mh;
   basis.mH = mH;
   basis.mA = mA;
   basis.mHp = mH;
   basis.sin_beta_minus_alpha = 1;
   basis.lambda_6 = 0;
   basis.lambda_7 = 0;
   basis.tan_beta = tb;
   basis.m122 = mH*mH/tb + (mh*mh - lambda1*v*v)/(tb*tb*tb); // Eq.(14)

   double amu = std::numeric_limits<double>::quiet_NaN();

   try {
      gm2calc::THDM model(basis, sm);

      amu = gm2calc::calculate_amu_1loop(model)
          + gm2calc::calculate_amu_2loop(model);
   } catch (const gm2calc::Error&) {}

   return amu;
}

int main()
{
   const double mA_start = 1, mA_stop = 100;
   const double tb_start = 1, tb_stop = 100;
   const int N_steps = 198;

   std::printf("%-20s%-20s%-20s%-20s\n",
               "# mA/GeV", "tan(beta)", "amu(II)", "amu(X)");

   for (int i = 0; i <= N_steps; i++) {
      for (int k = 0; k <= N_steps; k++) {
         const double tb = tb_start + i*(tb_stop - tb_start)/N_steps;
         const double mA = mA_start + k*(mA_stop - mA_start)/N_steps;

         const auto type_2 = calc_amu(mA, tb, gm2calc::thdm::Yukawa_type::type_2);
         const auto type_X = calc_amu(mA, tb, gm2calc::thdm::Yukawa_type::type_X);

         std::printf("%-20.10e%-20.10e%-20.10e%-20.10e\n",
                     mA, tb, type_2, type_X);
      }
   }

   return 0;
}
\end{lstlisting}
The function \lstinline|calc_amu| calculates $\amuBSM$ in the
\THDM\ at the two-loop level for a given value of $m_A$ and
$\tan\beta$ and a specified Yukawa type.  The remaining \THDM\ input
parameters in the mass basis are set to $m_h=126\GeV$,
$m_H=m_{H^\pm}=200\GeV$, $\sin(\beta-\alpha)=1$,
$\lambda_6=\lambda_7=0$,
$m_{12}^2=m_H^2/\tan\beta + (m_h^2 - \lambda_1 v^2)/\tan^3\beta$ and
$\lambda_1=\sqrt{4\pi}$.  In the \lstinline|main| function the loop
over $m_A$ and $\tan\beta$ is performed and $\amuBSM$ is
calculated for the type II and type X \THDM\ and the result is written
to the standard output.
The 2-dimensional output is shown in \figref{fig:mA-tb} for the two
types of the \THDM.

\subsection{Size of fermionic and bosonic contributions}
\label{sec:FB}

In the following we illustrate the calculation of the two-loop
fermionic and bosonic contributions, $\amuF$ and $\amuB$, separately.
For the illustration we perform a scan over $m_A$ for the demo
parameter scenario from \THDMC\ \cite{Eriksson:2009ws}, which is a
type II \THDM\ scenario where $m_H= 400\GeV$, $m_{H^\pm}=440\GeV$,
$\tan\beta=3$, $\sin(\beta-\alpha)=0.999$, $\lambda_6=\lambda_7=0$ and
$m_{12}^2=(200\GeV)^2$.  The following C source code shows the program
to perform the scan.
\begin{lstlisting}[language=c]
#include "gm2calc/gm2_1loop.h"
#include "gm2calc/gm2_2loop.h"
#include "gm2calc/gm2_error.h"
#include "gm2calc/THDM.h"

#include <stdio.h>

int main()
{
   const double mA_start = 130, mA_stop  = 500;
   const int N_steps = 200;

   printf("%-20s%-20s%-20s\n", "# mS/GeV", "amu(F)", "amu(B)");

   for (int i = 0; i <= N_steps; i++) {
      const gm2calc_THDM_mass_basis basis = {
         .yukawa_type = gm2calc_THDM_type_2,
         .mh = 125,
         .mH = 400,
         .mA = mA_start + i*(mA_stop - mA_start)/N_steps,
         .mHp = 440,
         .sin_beta_minus_alpha = 0.999,
         .lambda_6 = 0,
         .lambda_7 = 0,
         .tan_beta = 3,
         .m122 = 40000,
         .zeta_u = 0,
         .zeta_d = 0,
         .zeta_l = 0,
         .Delta_u = { {0,0,0}, {0,0,0}, {0,0,0} },
         .Delta_d = { {0,0,0}, {0,0,0}, {0,0,0} },
         .Delta_l = { {0,0,0}, {0,0,0}, {0,0,0} },
         .Pi_u = { {0,0,0}, {0,0,0}, {0,0,0} },
         .Pi_d = { {0,0,0}, {0,0,0}, {0,0,0} },
         .Pi_l = { {0,0,0}, {0,0,0}, {0,0,0} }
      };

      gm2calc_THDM* model = 0;
      gm2calc_error error = gm2calc_thdm_new_with_mass_basis(&model, &basis, 0, 0);

      if (error == gm2calc_NoError) {
         const double amuF = gm2calc_thdm_calculate_amu_2loop_fermionic(model);
         const double amuB = gm2calc_thdm_calculate_amu_2loop_bosonic(model);

         printf("%-20.10e%-20.10e%-20.10e\n", basis.mA, amuF, amuB);
      } else {
         printf("Error: %s\n", gm2calc_error_str(error));
      }

      gm2calc_thdm_free(model);
   }

   return 0;
}
\end{lstlisting}
The \SM\ input parameters and the configuration options are set to their
default values by passing 0 as the last two arguments to the function
\lstinline|gm2calc_thdm_new_with_mass_basis| that creates the \THDM\
model in line~39.  The individual bosonic and fermionic contributions
are calculated in lines~42--43 and written to the standard output in
line~45.
\begin{figure}
  \centering
  \includegraphics[width=0.49\textwidth]{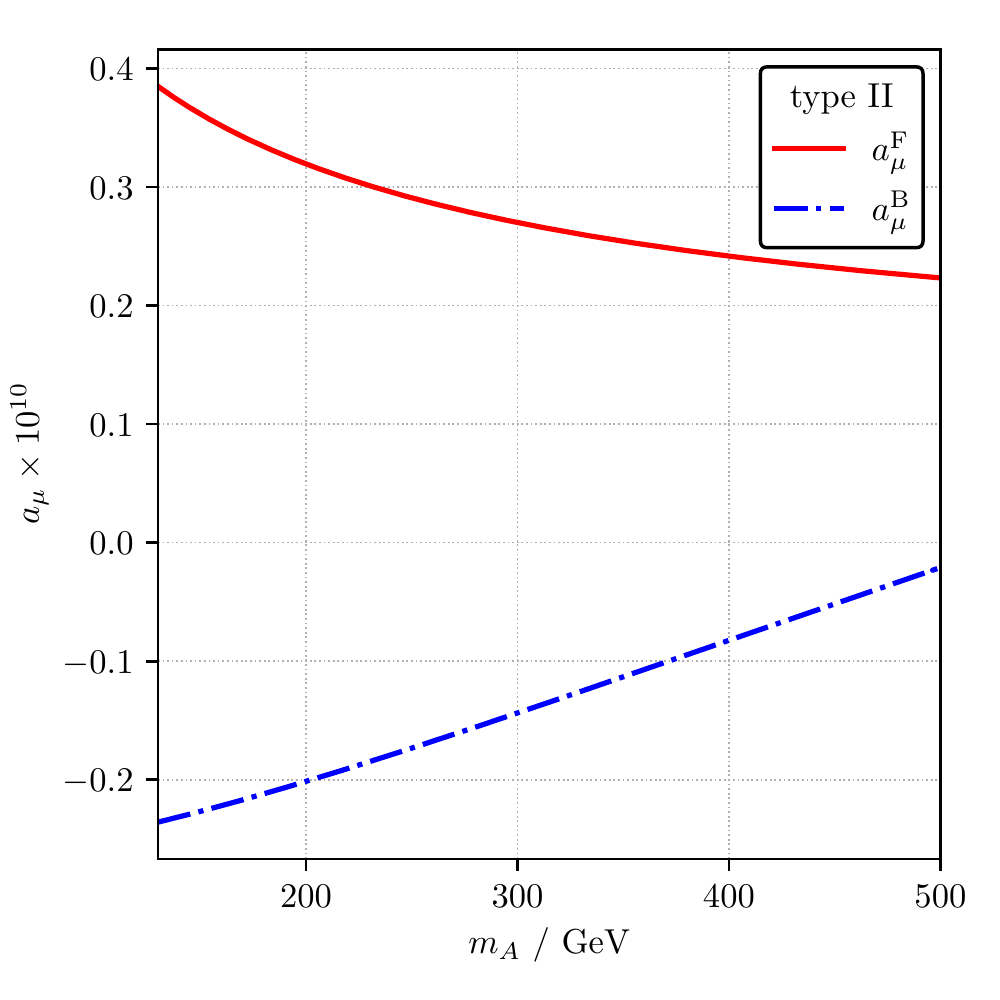}\hfill
  \includegraphics[width=0.49\textwidth]{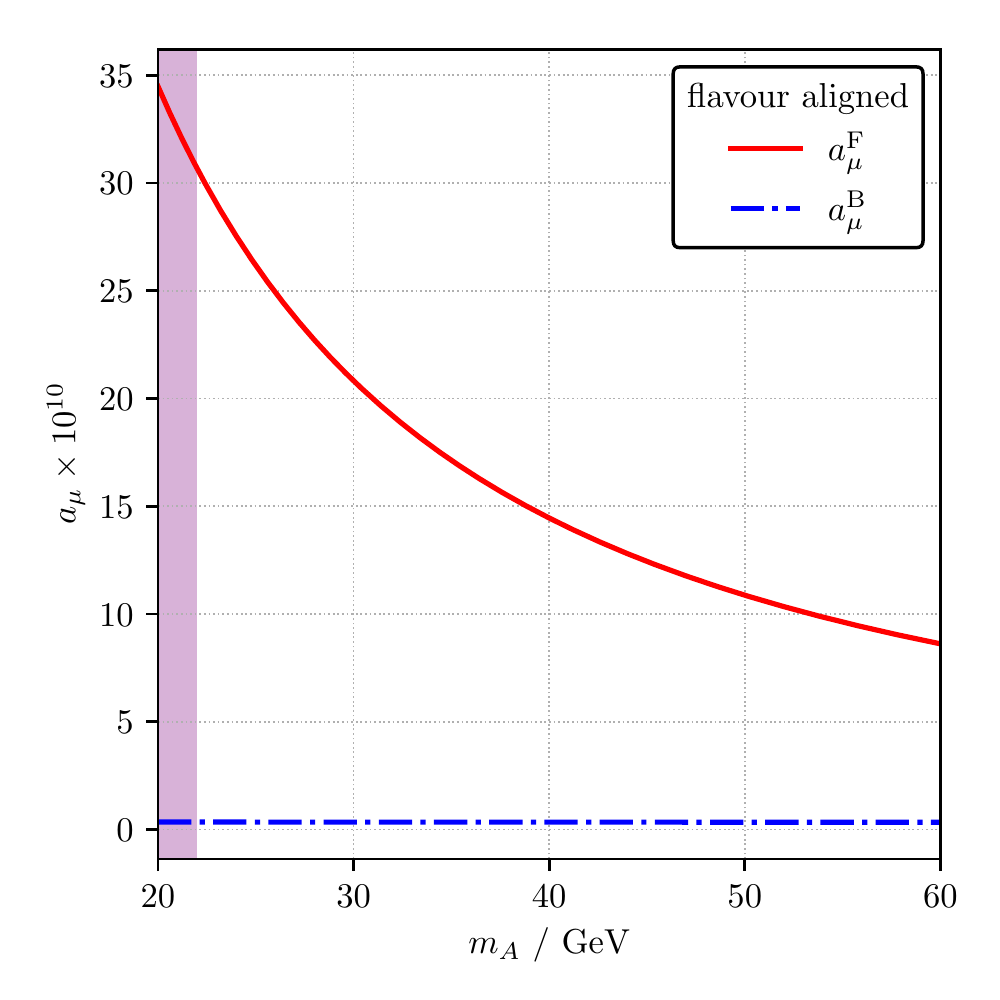}
  \includegraphics[width=0.49\textwidth]{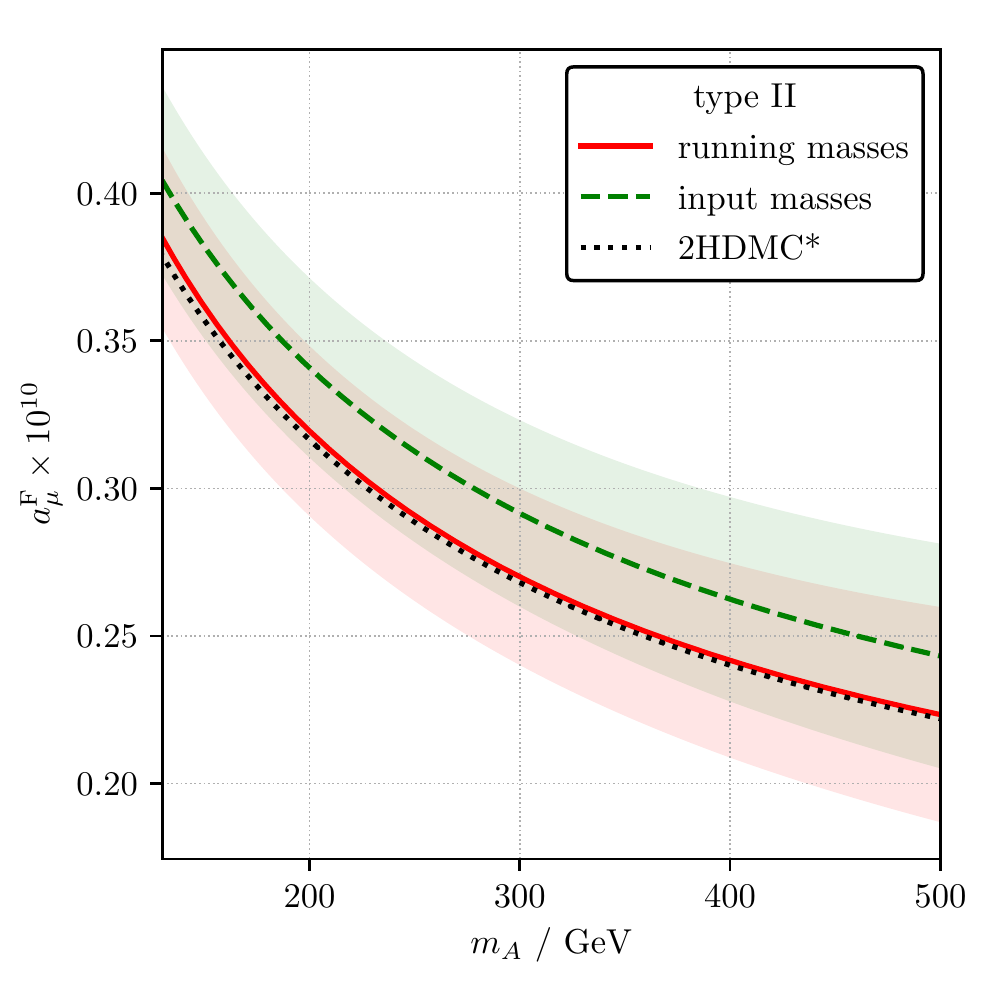}\hfill
  \includegraphics[width=0.49\textwidth]{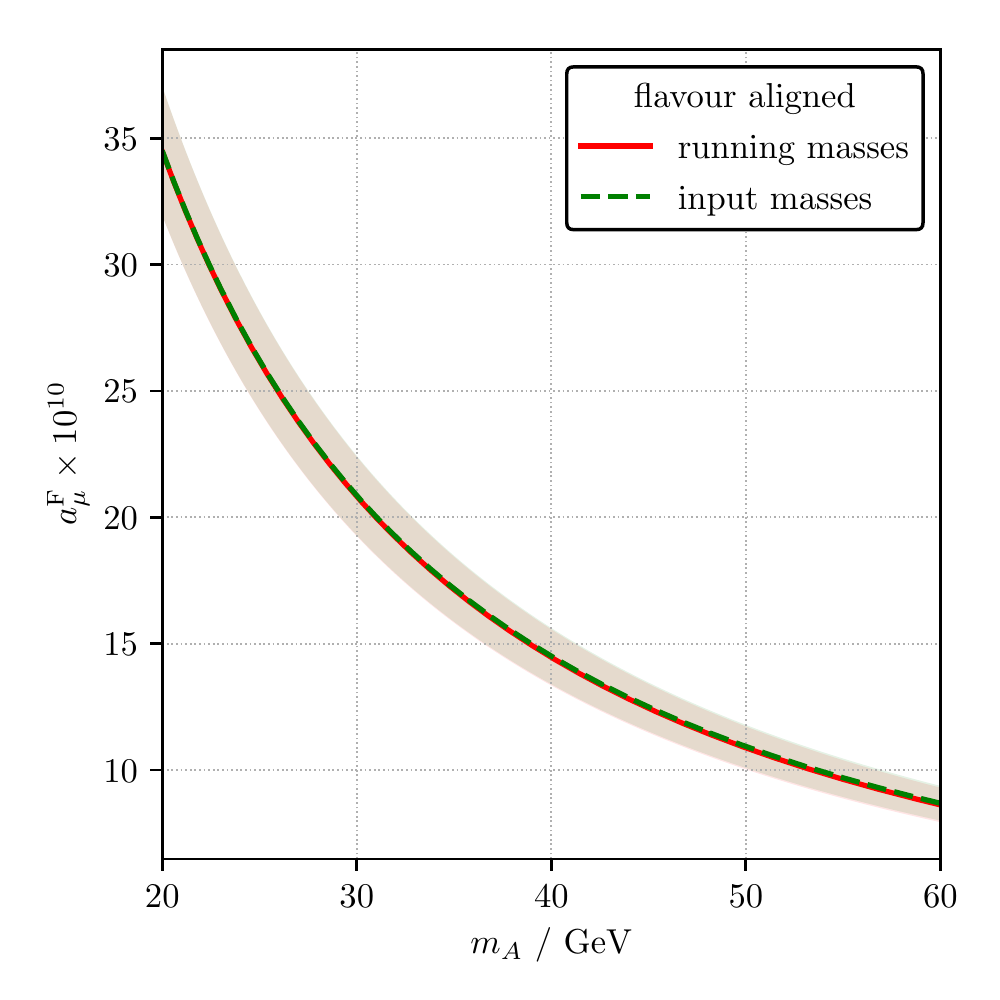}
  \caption{Different contributions to $\amuBSM$ in the \THDM. In the
    left panels we show scenarios from the type II \THDM\ as a
    function of $m_A$ with $m_H= 400\GeV, m_{H^\pm}=440\GeV$,
    $\tan\beta=3$, $\sin(\beta-\alpha)=0.999$, $\lambda_6=\lambda_7=0$
    and $m_{12}^2=(200\GeV)^2$.  In the right panels we show
    \FATHDM\ scenarios with $m_h = 125\GeV$, $m_H=150\GeV$,
    $m_{H^\pm}=150\GeV$, $\sin(\beta-\alpha)=0.999$, $\lambda_6=0$,
    $\lambda_7=0$, $\tan\beta = 2$, $m_{12}^2$ is fixed to avoid
    $h\rightarrow AA$ decays using a polynomial
    equation shown in the related source code, $\zeta_u=\zeta_d =
    -0.1$, $\zeta_l = 50$ based on scenarios found for \figurename~10
    of Ref.\ \cite{Cherchiglia:2017uwv}. In the top panels we show
    fermionic (red solid line) and bosonic (blue dashed-dotted line)
    two-loop contributions separately.  The top right panel also shows
    a purple region over the values of $m_A$ where it is possible to
    explain Eq.\ \eqref{eq:amuBSM}.  In the bottom panels we compare
    the fermionic two-loop contributions, when the running 3rd
    generation fermion masses shown in \eqref{eq:running_masses} are
    used (red solid line) and when the 3rd generation input masses
    from \eqref{eq:input_masses} are used (green dashed line).  The
    red and green bands show the corresponding uncertainties.  The
    bottom left panel also shows two-loop fermionic contributions
    calculated from \THDMC\ as a black dotted line.  
    *This is not a vanilla \THDMC\ calculation, the
    \SM-Higgs-like two-loop contributions have been removed, and the two-loop 
    charged Higgs Barr-Zee contributions have been added as explained in the text.  }
  \label{fig:FB}
\end{figure}
In the top left panel of \figref{fig:FB} we show the results for the
fermionic (red solid line) and bosonic (blue dashed-dotted line)
contributions as a function of $m_A$ for this demo scenario from
\THDMC.  For the shown parameter space the fermionic contributions
decrease, while the bosonic contributions increase for increasing
$m_A$.  Furthermore, the bosonic contributions are negative, while the
fermionic contributions are positive, which leads to a partial
cancellation of the two contributions.  However, as is generally the
case for the type II \THDM\ scenarios that are not already excluded by
other constraints, the scenarios we plot here cannot explain the large
deviation between the Standard Model prediction and experiment given
in Eq.\ \eqref{eq:amuBSM}.

If instead we consider the \FATHDM\ scenarios, as in the top right panel
of \figref{fig:FB} it is now possible to explain large deviations.
Here we fix the following parameters $m_h = 125\GeV$, $m_H=150\GeV$,
$m_{H^\pm}=150\GeV$, $\sin(\beta-\alpha)=0.999$, $\lambda_6=0$,
$\lambda_7=0$, $\tan\beta = 2$, $\zeta_u=\zeta_d=-0.1$,
$\zeta_l = 50$ based on results used for \figurename~10
of Ref.\ \cite{Cherchiglia:2017uwv}. It should be noticed that,
  once the masses for the charged and CP-even scalar are close
  together and below around 300 GeV, electroweak as well as unitarity
  and perturbativity constraints can be evaded for arbitrarily low
  values of $m_A$ \cite{Broggio:2014mna}. The input parameters 
  here are chosen accordingly.
Since $m_A < m_{h_\SM}/2$, it is possible for $h \rightarrow A A$ decays to occur
unless we enforce the coupling $C_{hAA}=0$.\footnote{The coupling
    $C_{hAA}$ has been discussed in detail 
    in Ref.\ \cite{Cherchiglia:2017uwv}. 
    Setting $C_{hAA}=0$ directly corresponds to the choice $\Lambda_6=0$, where
    $\Lambda_6$ is a potential parameter in the so-called Higgs basis,
    and to a specific relation between all potential parameters in the
    general basis. It has been checked that
    although non-null values for $C_{hAA}$ can be allowed, they are
    experimentally strongly constrained. Thus, for simplicity, we have
    adopted $C_{hAA}=0$ in our analysis, which automatically
    guarantees that $\lambda_1$ (or equivalently $m_{12}^2$) will not
    be chosen in a region excluded by experimental constraints or by
    unitarity or perturbativity.}  This fixes the value of $\lambda_1$ 
according to Eq.\ (12) in Ref.\ \cite{Cherchiglia:2017uwv}, 
which can be set in the mass basis using $m_{12}^2$ and applying the relations in
Eqs.\ (2.12)--(2.13) in Ref.\ \cite{Cherchiglia:2016eui}.  
This leads to the fitted 2nd-order polynomial relation with a dependence on $m_A$ seen in the source code below.  

These scenarios have a very light pseudoscalar mass, but LHC limits
are much weaker compared to the type II case and can be evaded for
these scenarios.  The two-loop fermion contributions rise rapidly as
the pseudoscalar mass decreases, dominating over the two-loop bosonic
contributions, though the latter are just large enough to have an
impact on constraints from $\amu$.  Note that for higher values of
$m_A$ it is possible to get larger bosonic contributions as can be
seen in \figurename~10 of Ref.\ \cite{Cherchiglia:2017uwv}.  In the
scenarios we plot here the one-loop contributions, which are not shown
in \figref{fig:FB}, have a negative effect on the contributions, with
a size of approximately one-third of the two-loop fermionic
contributions.  Thus it can be seen that $\amuBSM$ can be explained
with a very low $m_A$ for the values of $m_A$ in the purple region.
The scan for this scenario can be performed with the following C
source code:
\begin{lstlisting}[language=c]
#include "gm2calc/gm2_1loop.h"
#include "gm2calc/gm2_2loop.h"
#include "gm2calc/gm2_error.h"
#include "gm2calc/THDM.h"

#include <stdio.h>
#include <math.h>

int main()
{
   const double mA_start = 20, mA_stop = 60;
   const int N_steps = 200;

   printf("%-20s%-20s%-20s%-20s\n", "# mS/GeV", "amu(F)", "amu(B)", "amu(1)");

   for (int i = 0; i <= N_steps; i++) {
      const double mA = mA_start + i*(mA_stop - mA_start)/N_steps;

      const gm2calc_THDM_mass_basis basis = {
         .yukawa_type = gm2calc_THDM_aligned,
         .mh = 125,
         .mH = 150,
         .mA = mA,
         .mHp = 150,
         .sin_beta_minus_alpha = 0.999,
         .lambda_6 = 0,
         .lambda_7 = 0,
         .tan_beta = 2,
         // Polynomial fit following Eq. (2.12) from PRD
         .m122 = 3187.3 + mA*(3.27803 + 0.0165557*mA),
         .zeta_u = -0.1,
         .zeta_d = -0.1,
         .zeta_l = 50,
         .Delta_u = { {0,0,0}, {0,0,0}, {0,0,0} },
         .Delta_d = { {0,0,0}, {0,0,0}, {0,0,0} },
         .Delta_l = { {0,0,0}, {0,0,0}, {0,0,0} },
         .Pi_u = { {0,0,0}, {0,0,0}, {0,0,0} },
         .Pi_d = { {0,0,0}, {0,0,0}, {0,0,0} },
         .Pi_l = { {0,0,0}, {0,0,0}, {0,0,0} }
      };

      gm2calc_THDM* model = 0;
      gm2calc_error error = gm2calc_thdm_new_with_mass_basis(&model, &basis, 0, 0);

      if (error == gm2calc_NoError) {
         const double amu1 = gm2calc_thdm_calculate_amu_1loop(model);
         const double amuF = gm2calc_thdm_calculate_amu_2loop_fermionic(model);
         const double amuB = gm2calc_thdm_calculate_amu_2loop_bosonic(model);

         printf("%-20.10e%-20.10e%-20.10e%-20.10e\n", basis.mA, amuF, amuB, amu1);
      } else {
         printf("Error: %s\n", gm2calc_error_str(error));
      }

      gm2calc_thdm_free(model);
   }

   return 0;
}
\end{lstlisting}
%

\subsection{Running fermion masses}

In this subsection we study the effect of using the input vs.\ running
fermion masses in the two-loop fermionic contributions as described in
\secref{sec:running_couplings}, and compare the results with \THDMC.  
The following \mathematica\ source
code shows a program to perform a scan over $m_A$ using the same type II \THDM\
parameter region as in \secref{sec:FB}.
\begin{lstlisting}[language=mathematica]
Install["bin/gm2calc.mx"];

CalcAmu[mA_] :=
    {amu2LF, Damu} /. GM2CalcAmuTHDMMassBasis[
        yukawaType        -> 2,
        Mhh               -> { 125, 400 },
        MAh               -> mA,
        MHp               -> 440,
        sinBetaMinusAlpha -> 0.999,
        lambda6           -> 0,
        lambda7           -> 0,
        TB                -> 3,
        m122              -> 200^2
    ];

(* mA values in [130,500] GeV *)
mAValues = Subdivide[130, 500, 200];

(* calculation w/o running couplings *)
GM2CalcSetFlags[runningCouplings -> False];
noRunning = CalcAmu /@ mAValues;

(* calculation w/ running couplings *)
GM2CalcSetFlags[runningCouplings -> True];
withRunning = CalcAmu /@ mAValues;

data = { mAValues,
         noRunning[[All,1]], noRunning[[All,2]],
         withRunning[[All,1]], withRunning[[All,2]] };

Export["running.txt", N @ Transpose @ data, "Table"];
\end{lstlisting}
The function \lstinline|CalcAmu| calculates the two-loop fermionic
contribution $\amuF$ and the uncertainty $\Damu$ for a given value of
$m_A$ and the mass basis input parameters defined above.  In
line~20 the usage of running fermion masses is disabled and the
calculation is performed in the subsequent line.  Similarly, in
line~24 the usage of running fermion masses is enabled and the
calculation is performed in the subsequent line.  The results are
collected in the variable \lstinline|data| and are exported to a file
in line~31.  We also used a very similar script to do the same
calculations for the \FATHDM\ scenarios discussed in \secref{sec:FB}.

The effect of using running fermion masses in the two-loop fermionic
contributions is shown in \figref{fig:FB} for the type II (bottom left
panel) and flavour aligned (bottom right panel) scenarios matching
those described in the previous section.  The red solid line shows the
value of $\amuF$ when the running masses \eqref{eq:running_masses} are
used, i.e.\ the 3rd generation fermion masses are run to the scale of
the Higgs boson in the two-loop fermionic Barr-Zee Feynman diagrams.
Note that although the vertical axes are slightly different,
the red lines shown in the bottom panels are identical to the red
lines from the corresponding panels immediately above them, which
were discussed in the previous section. The green dashed lines show
the value of $\amuF$ when input fermion masses listed in
\eqref{eq:input_masses} are used.  In both scenarios that we look at
the value of $\amuF$ is smaller when running masses are used,
though the difference is only distinguishable for the type II case where the
size of the contributions is much smaller.  The reason for this is
that due to the negative fermion mass $\beta$ functions the running
masses are numerically smaller than the corresponding input masses in
the shown scenarios, which leads to a systematic reduction of the
fermionic two-loop contributions.

In addition to the red solid and green dashed lines from \GMTCalc, as
the scenario in question is a benchmark point for
\THDMC\ \cite{Eriksson:2009ws}, we show in the bottom left panel of
\figref{fig:FB} the corresponding result obtained with \THDMC\ 1.8.0
as black dotted line.  Note that at the two-loop level
\THDMC\ includes only fermionic contributions to $\amuBSM$.
Furthermore, \THDMC\ does not subtract the
contributions from the \SM\ Higgs boson and does not include the
two-loop contributions from the charged Higgs boson.  Therefore to
obtain this black dotted line we have thus subtracted the
two-loop \SM\ Higgs contributions from the \THDMC\ result and added
the two-loop contributions from the charged Higgs boson.  Since
\THDMC\ inserts running fermion masses into the fermionic
contributions, the black dotted line can be compared to the red solid
line in the figure.  There is a small deviation between these two
lines, which originates from the inclusion of fermionic Barr-Zee
diagrams with an internal $Z$ boson in \GMTCalc, which are not
included in \THDMC.

In the bottom panels of \figref{fig:FB} we also show the uncertainties
calculated with \eqref{eq:damu} as lighter shaded regions of the
corresponding color about the red and green lines. In the bottom left
panel the red and green lines both lie within the uncertainty estimate
for the alternative prediction (shaded green and shaded red regions
respectively) for all values of $m_A$ plotted.  This is also true in
the bottom right panel though there is no visible distinction between
the red and green lines or their uncertainties here. Since the
difference between the lines is of higher order, this indicates that
our uncertainty estimate is working as expected and accounts for the
expected higher order corrections.

\section{Summary}

We have presented version 2 of \GMTCalc, with its new capability to
calculate the \BSM\ contributions to the anomalous magnetic moment of
the muon in the \THDM.  The contributions include all the one-loop
diagrams, two-loop fermionic Bar-Zee diagrams, as well as the bosonic
two-loop contributions. The new version of \GMTCalc\ provides the
calculation of the \THDM\ contributions with a precision of up to
$\order(m_\mu^4)$ at one-loop and $\order(m_\mu^2)$ at two-loop level
along with an estimate of uncertainty of the evaluation.
\GMTCalc\ performs this state of the art precision calculation at high
speed, with execution times that can be as short as
$\order(0.05\unit{ms})$ per point, allowing for rapid sampling of the
parameter space.

\GMTCalc\ is easy to configure and run.  The user can select well-known
types of the \THDM, specifically type I, II, X, Y as well as the
flavour-aligned version (\FATHDM), or the fully general \THDM.  For the
latter the user can specify the inputs as deviations away from the
flavour alignment of the \FATHDM\ (or from type I, II, X, and Y) or by
directly specifying the more fundamental Yukawa matrices, $\Pi_f$
defined in Eq.\ \eqref{eq:Piprime}.  The user can also decide whether
they will give inputs in the gauge basis using $\lambda_{1,\ldots,7}$
or the mass basis using $m_{h,H,A,H^\pm}$, $\lambda_{6,7}$, and
$\sin(\beta-\alpha)$.  The input parameters and settings can be
specified in an SLHA-like input file, mirroring the original
\MSSM\ version.  Additionally, \GMTCalc\ can be interfaced to other
programs using C++, C, \mathematica, or \python, the latter being a new interface developed for \GMTCalc~2.

For each of these interfaces we presented simple and easy to follow
usage examples in \secref{sec:implementation}, that are
straightforward to adapt.  In \secref{sec:applications} we have
also presented some applications and results, demonstrating different
features of the code.  In each of these we show the source code in the
manual and also provide them as supplementary
files.  \GMTCalc\ is actively developed on GitHub
and users with any questions may contact the authors through our
\href{https://github.com/GM2Calc/GM2Calc}{GitHub page} or directly by
email.

\section*{Acknowledgments}

We would like to thank the other authors of version 1 of \GMTCalc, M. Bach, H. G. Fargnoli, C. Gnendiger, R. Greifenhagen, J. Park, and S. Pa{\ss}ehr.  We also thank T Gonzalo and C. S. Fonseca for their assistance with work dealing with two-loop contributions to the anomalous magnetic moment of the muon.
The research placement of D.J. for this work was supported by the Australian Government Research Training Program (RTP) Scholarship and the Deutscher Akademischer Austauschdienst (DAAD) One-Year Research Grant.  The work of P.A. was supported by the Australian Research Council Future Fellowship grant FT160100274.  The work of P.A. and C.B. was also supported with the Australian Research Council Discovery Project grant DP180102209.
This project was undertaken with the assistance of resources and services from the National Computational Infrastructure, which is supported by the Australian Government.  We thank Astronomy Australia Limited for financial support of computing resources.

\addcontentsline{toc}{section}{References}

\bibliography{paper,TheoryWPbiblio,MuonMoment2HDM}
\bibliographystyle{JHEP}
\end{document}